\documentclass[aps,pre,twocolumn,superscriptaddress,longbibliography]{revtex4-2}

\usepackage{soul}

\usepackage{amsmath}
\usepackage{amssymb}
\usepackage{graphicx}
\usepackage{float}
\usepackage{siunitx}
\usepackage{xcolor}
\usepackage{cancel}
\usepackage{mathtools,lipsum, nccmath}

\newcommand{\VEC}[1] {\mathbf{#1}}
\newcommand{\HAT}[1] {\hat{\mathbf{#1}}}
\newcommand{\n}{\VEC{\hat{n}}}
\newcommand{\ket}[1]{\vert #1 \rangle}
\newcommand{\bra}[1]{\langle #1 \vert}
\newcommand{\braket}[2]{\langle #1 \vert #2 \rangle }
\newcommand{\tum}[2]{\mathcal{T}_{#1}[#2]}
\newcommand{\pdv}[2]{\frac{\partial #1}{\partial #2}}

\begin{document}

\title{Bacterial chemotaxis considering memory effects}

\author{Manuel Mayo}
\affiliation{F\'isica Te\'orica, Universidad de Sevilla, Apartado de Correos 1065, E-41080, Sevilla, Spain}
\affiliation{Departamento de F\'isica, Facultad de Ciencias F\'isicas y Matem\'aticas, Universidad de Chile, Avenida Blanco Encalada 2008, Santiago, Chile}

\author{Rodrigo Soto}
\affiliation{Departamento de F\'isica, Facultad de Ciencias F\'isicas y Matem\'aticas, Universidad de Chile, Avenida Blanco Encalada 2008, Santiago, Chile}

\date{\today}

\begin{abstract}
Bacterial chemotaxis for the case of  \textit{E.\ coli} is controlled by methylation of chemoreceptors, which in a biochemical pathway regulates the concentration of the CheY-P protein that finally controls the tumbling rate. As a consequence, the tumbling rate adjusts to changes in  the concentration of relevant chemicals, such as to produce a biased random walk toward  chemoattractants of against the repellers.
Methylation is a slow process, implying that the internal concentration of CheY-P is  not instantaneously adapted to the environment, and therefore the tumbling rate presents a memory.
This implies that the Keller--Segel equations used to describe chemotaxis at the macroscopic scale, which assume a local relation between the bacterial flux and the chemical gradient, cannot be fully valid as memory and the associated nonlocal response are not taken into account.
To derive the new equations that replace the Keller--Segel ones, we use a kinetic approach, in which a kinetic equation for the bacterial transport is written considering the dynamics of the protein concentration. When memory is large, the protein concentration field must be considered a relevant variable on equal foot as the bacterial density. Working out in detail the Chapman--Enskog method,  the dynamical equations for these fields are obtained, which have the form of reaction-diffusion equations with flux and source terms depending on the gradients on the chemical signal. Also, the transport coefficients are obtained entirely in terms o the microscopic dynamics, showing important symmetry properties and giving their values of the case of  \textit{E.\ coli}. Solving the equations for an inhomogeneous signal it is shown that the response is nonlocal, with a smoothing length as large as \SI{170}{\micro\meter} for \textit{E.\ coli}. The homogeneous response and the relaxational dynamics are also studied in detail. For completeness, the case of small memory is also studied, in which case the Chapman--Enskog method reproduces the Keller--Segel equations, with explicit expressions for the transport coefficients.

\end{abstract}

\maketitle


\section{Introduction}
The pioneering works of Berg and coworkers gave the study of bacterial transport a quantitative character, with detailed mathematical models~\cite{berg1972chemotaxis,berg2008coli}. 
The most relevant transport process is chemotaxis, where bacteria move along or against chemical gradients. 
The current understanding of chemotaxis of flagellated bacteria like \textit{E.\ coli} is that they perform a biased random walk, with longer walks in the direction of the attractant and shorter in the opposite direction. For \textit{E.\ coli}, this is achieved by modulating the tumble rate in response to changes in the chemical concentration as they move. As the tumbling process is stochastic \cite{berg1972chemotaxis,korobkova2004molecular,tu2005white}, this description immediately implies that bacterial diffusion appears together with chemotaxis.

At the macroscopic scale, bacterial chemotaxis is described by the Keller--Segel equations, which couple the dynamics of the bacterial density $\rho$ to the ligand concentration $l$ (food, chemoattractant, or chemorepellent)~\cite{keller1970,keller1971model}.  In presence of a ligand gradient,  the bacterial flux has a chemotactic term, proportional to the ligand gradient, which is added to the diffusive flux. Hence, for the density one gets
\begin{subequations}\label{eqs.ks}
\begin{align} 
\pdv{\rho}{t} &= -\nabla\cdot \VEC J, \label{eq.ks1}\\
\intertext{with}
\VEC J &= -D\nabla \rho + \mu \rho \nabla l. \label{eq.ks2}
\end{align}
\end{subequations}
In case cell division or death processes also take place, a source term can be added to the first equation. Here, $D$ is the diffusion coefficient and $\mu$ the chemotactic mobility, which is positive for chemoattractants and negative for repellers.
Coupled to this, the Keller--Segel system is complemented by an equation for the ligand concentration, which is a diffusion equation with a sink term representing the depletion of the ligand by bacteria. In this article, we will analyze the equation for the bacterial density and, therefore, we will assume that the ligand gradient is imposed.
Eq.~\eqref{eq.ks2} uses a linear coupling with the ligand, and more refined relations have been obtained. For example, the flux being a nonlinear function of $\nabla l$ or the mobility depending on the absolute value of the ligand \cite{rivero1989transport,ford1992relationship,tu2008modeling,tu2013quantitative,cremer2019chemotaxis}.

The Keller--Segel equations can be derived from the run-and-tumble dynamics applying different techniques of nonequilibrium statistical mechanics \cite{alt1980biased,rivero1989transport,ford1992relationship,schnitzer1993theory,chen1998perturbation,bearon2000modelling,kasyap2014instability}. For that it is assumed that  gradients of the different fields are  small compared to the  scale of the microscopic agents (bacteria in this case). That is, the Keller--Segel equations result from a process of coarse graining to the so-called hydrodynamic scale. This process does not only generate the dynamical equations, but also the transport coefficients $D$ and $\mu$.

The tumbling rate in bacteria like \textit{E.\ coli} is controlled by the concentration of the CheY-P protein inside the bacterial body and chemotaxis results from varying the equilibrium concentration of this protein as response to changes in the ligand concentration~\cite{tu2005white,tu2008modeling,tostevin2009mutual,lan2012energy,Ito2015model}. The change of Chey-P concentration is a result of a biochemical pathway, where an important element in the methylation of the chemoreceptors. Methylation is a slow process and therefore the  concentration of CheY-P presents a long memory time, introducing a new temporal scale of tens of seconds \cite{yuan2012adaptation,tu2013quantitative,zhang2018motor,figueroa20203d} and an associated length scale of hundreds of micrometers \cite{figueroa2020coli,villa2023kinetic}. These scales are comparable to those appearing in microfluidic devices and in the complex natural environments where bacteria live, such as soils, organs, or pores (see, for example, \cite{sarkar1994transport,mao2003sensitive,ford2007role,licata2016diffusion,bhattacharjee2019bacterial}). Also, in sea water, microscale nutrient patches appear in the form of chemical pulses~\cite{blackburn1998microscale,stocker2012marine}. 
As a consequence, the simple Keller--Segel equations are not expected to be valid on those scales and new macroscopic equations valid for those scales need to be derived. 
One possible approach could be to include terms of higher gradient order in the chemotactic flux or the use of nonlocal kernels in the flux. This has been the approach followed by some authors.  In Ref.~\cite{Si2012} the one-dimensional case is considered, which is extended into two dimensions at the expense of performing some uncontrolled approximations~\cite{Si2014}. Similarly, in Ref.~\cite{Chalub2006} the need of introducing additional variables make that the resulting macroscopic equations are not fully based on the microscopic model. In Refs.~\cite{XueandOthmer2009,Xue2015}, although starting from a kinetic model with memory the macroscopic equations are obtained in the long time limit, in which case the Keller-Segel equations are recovered. Finally, it must be noted that in those works, the fluctuations of the internal variables are completely neglected, which we show below is not correct, at least for the case of \textit{E.\ coli}.

Here, we adopt a different strategy to obtain the macroscopic equations. First, we show that the bacterial concentration of CheY-P, $\rho_X$, is a relevant field, as important to the density, to describe the chemotactic process. Hence, it is necessary to derive the coupled equations for $\rho$ and $\rho_X$, which is the main purpose of this article. 
As different bacteria strains can differ on their tumbling and methylation memory times, the latter not being necessarily large as in the case of \textit{E.\ coli}, we will derive the hydrodynamic equations for both for cases of large and small memory.

The article is organized as follows. In Sect.~\ref{sec.kt} we first present the mathematical model for chemotaxis  of bacteria like \textit{E.\ coli} at the individual scale, which is incorporated into a kinetic theory  for an ensemble of bacteria in presence of chemical signals.  We end this section presenting some relevant mathematical properties of the kinetic equation. In Sect.~\ref{sec.ChE1} we apply the Chapman-Enskog procedure, which is a systematic approach to derive coarse-grained equations for successive temporal scales, to the kinetic equation  when  the memory is not necessarily large. In this case, the usual Keller--Segel model is recovered. The Chapman-Enskog method for the case of large memory is worked out in Sect.~\ref{sec.ChE2}, where new equations are obtained together with the associated transport coefficients. This results in Eqs.~(\ref{complet_rho_eq}-\ref{eqs.constlaws}) that replace the Keller--Segel  equations \eqref{eqs.ks} and constitute the principal contribution of this article.
Several relevant cases are studied in detail in Sect.~\ref{sec.analysis}. The short section \ref{sec.ecoli} gives the values of the transport coefficients for the case of \textit{E.\ coli}. Finally, conclusions are given in Sect.~\ref{sec.conclusions}.

\section{Kinetic Theory of Bacterial Chemotaxis}\label{sec.kt}

\subsection{Chemotactic model}

Bacteria like \textit{E.\ coli} move in fluids following the so-called run-and-tumble dynamics~\cite{berg1972chemotaxis}. In the run phase, bacteria move with roughly constant speed and direction, process that is interrupted by rapid reorientations called tumbles. The latter are initiated when there is a reversal in the direction of rotation (from counterclockwise, CCW, to clockwise, CW) of one or multiple flagella~\cite{berg2008coli,taktikos2013motility}. This process of motor switch is governed by the concentration $y(t)$ of the phosphorylated CheY protein (CheY-P)  inside the bacterial body.  In a simple description, based on the biochemistry of the molecular motor, Tu and Grinstein proposed a model to describe the tumbling process as a two-state activated system~\cite{tu2005white}. In this model, the activation free energy barrier $\Delta G$ and, hence, the transition rate $\nu\sim e^{-\Delta G/kT}$ from the CCW to the CW state depend on the instantaneous concentration $y(t)$. Considering a Taylor expansion of $\Delta G$ and defining the normalized concentration deviation, $X(t ) = [y (t ) - \langle y\rangle]/\sigma_y$, it results
$\nu=\nu_0 e^{\lambda X}$,    
where $\sigma_y$ is the standard deviation of $y$ and the dimensionless parameter $\lambda$ measures the sensitivity of the tumbling rate to changes in $X$. In the Tu and Grinstein model, $X$ is a Gaussian variable of null average and unit variance, with correlation time $\tau$. In summary, $X$ is described therefore as an Ornstein--Uhlenbeck process. By tracking \textit{E.\ coli} (RP437 bacteria in motility buffer supplemented with serine), this model was validated and allowed to extract the model parameters: $\nu_0=\SI{0.22}{\second^{-1}}$, $\tau=\SI{19}{\second}$, and $\lambda=1.62$~\cite{figueroa20203d}. As advanced, the memory time $\tau$, which is governed by the methylation process, is long compared to the mean run time. Associated to it, there is a characteristic memory length $L=V\tau$, where $V$ is the bacterial swim speed. Using $V = \SI{27}{\micro\meter/\second}$ from Ref.~\cite{figueroa20203d}, gives $L=\SI{500}{\micro\meter}$. This relatively large correlation length, comparable to pore sizes in natural environments, generate nonlocal responses that we aim to consider in our description. Also, the  large value of $\lambda$ indicates that the fluctuations in $X$ are important and their effects cannot be disregarded.

Bacteria respond to chemotactic signals by modulating their tumbling rate. As these microorganisms are too small to measure gradients along their body, they integrate the ligand signal on their run, with a chemical pathway that can be modeled by  three principal components: methylation level $m(t)$, kinase activity $a(t)$, and the aforementioned $y(t)$. With these elements, the tumbling rate modulation can be summarized as follows. Starting from a steady state with $a=y=0$ and $m$ adapted to the ligand concentration $l$, an abrupt change of $l$ triggers a rapid decrease of  $y$ and $a$ to negative values. Later, on a larger methylation time scale $\tau_m$ (of the same order of $\tau$) $m$ adapts to a new value, and $a$ and $y$ become null again. Hence, for a time $\tau_m$, $y$ becomes negative as a response to changes on $l$. That is, $y$ follows the negative of the ligand time derivative. The dynamics of this process can be expressed mathematically using coupled Langevin equations for $a(t) $, $y(t)$, and $m(t)$~\cite{tu2008modeling,tostevin2009mutual,lan2012energy,Ito2015model}. Theoretical analysis of these equation have studied the effects of memory on the response when time-dependent signals are imposed~\cite{dufour2014limits,Ito2015model,wong2016role}.
In Ref.~\cite{villa2023kinetic}, by adiabatically eliminating the fast modes of these equations, and keeping linear couplings, we showed that the chemotactic coupling can be described by a modification of the Tu and Grinstein dynamical equation for $X$, to incorporate the coupling with the ligand as
\begin{align}\label{langevin_eq_lineal}
    \dot{X}=-\frac{X+b \dot{l}}{\tau}+\sqrt{\frac{2}{\tau}} \xi,
\end{align}
where in this Langevin equation $\xi$ is a white noise of correlation $\langle\xi(t) \xi(t')\rangle = \delta(t-t')$, $b$ is the coupling constant to the ligand  (positive for attractants and negative for repellers), and the square root prefactor guarantees that if $\dot l$ is constant, $X$ reaches a normal distribution of unit variance.

This simple model has been improved in several aspects. Motor adaptation implies that the $\nu$ dependence on $X$ is not exponential~\cite{cluzel2000ultrasensitive,bai2010conformational,yuan2012adaptation,zhang2018motor}. Also, the fluctuation of CheY-P are not small, implying that the Langevin equation for $X$ needs to include non-linear coupling terms~\cite{colin2017multiple,keegstra2017phenotypic}. There are several models, still under study, and therefore we consider the  most general description for the stochastic dynamics of $X$. In principle the noise intensity can also depend on $X$. With a change of variable it is always possible to make this intensity constant, resulting in the general equation
\begin{align}\label{langevin_eq_general}
    \dot{X}=-\frac{A(X,l)+B(X,l) \dot{l}}{\tau}+\sqrt{\frac{2}{\tau}} \xi.
\end{align}
Here, due to the possible change of variable, $X$ is not exactly the normalized CheY-P concentration, but it is closely related to it. To avoid overloading the language, we will continue calling it the  normalized CheY-P concentration. $A$ and $B$ are dimensionless functions of order one, and we keep $\tau$ as the only relevant time scale. We assume that in absence of noise, for any value of $l$, there is a single stable fixed point.  Also, to ensure the existence of a linear response regime, the scalings $A(X)\sim X$ and $B(X)\sim b$ should be imposed for small $X$.  And for the tumbling rate we take, 
\begin{align} \label{tumblingrate}
\nu=\nu_0 C(X,l),
\end{align}
where $C$ is a monotonous increasing function of $X$, normalized such that $C(0)=1$. In what follows, we will work with this general model, but for concrete results we will use the linear model, corresponding to 
\begin{align} \label{linmodel}
A_\text{lin}(X)&=X, &B_\text{lin}(X)&=b, & C_\text{lin}(X)&=e^{\lambda X}.
\end{align}
Finally, Eq.~\eqref{langevin_eq_general} is coupled to the bacterial motion because $\dot l$ is the rate of change of $l$ in the comoving frame of the swimmer. For a bacterium moving with speed $V$ along the director $\hat{\mathbf{n}}$, it is written as the Lagrangian derivative
\begin{align}\label{lagrang_derivative_ligand}
    \Dot{l}=V \hat{\mathbf{n}} \cdot \nabla l+\frac{\partial l}{\partial t}.
\end{align}

In this representation, chemotaxis is rationalized as follows. For simplicity, we consider the linear model, but the analysis is analogous for the general case. For a swimmer moving parallel to a chemoattractant gradient ($\hat{\mathbf{n}} \cdot \nabla l>0$ and $b>0$), the normalized protein $X$ becomes negative on average (that is, $y$ becomes smaller than $\langle y\rangle$), reducing by Eq.~\eqref{tumblingrate} the tumbling rate, and exactly the opposite results for a swimmer moving against the gradient. The result is a biased run-and-tumble random walk, with longer runs in the direction of the gradient. Similarly, for a chemorepellent ($b<0$), the motion is biased against the gradient.
In absence of memory and fluctuations, $X$ adapts instantaneously the ligand rate of change, $X=-b\dot l$, resulting in the tumbling rate $\nu=\nu_0 e^{-\lambda b \dot l}$. This expression or similar ones have been used to describe chemotaxis at the kinetic level~\cite{Schnitzer1993,Chen2003,Lushi2012,Kasyap2012,kasyap2014instability,Bearon2000,tindall2008}.
Here, we go beyond this approximation, considering the effects of fluctuations and memory in the chemotactic process. 

In Ref.~\cite{villa2023kinetic} we proposed a kinetic equation that incorporated all the elements of the linear model, which we solved to study the stationary chemotactic mobility and the linear response to signals varying in space and time. Importantly, the obtained response is nonlocal in space and time, as an effect of the memory time. 
The method to build the solutions is not easy to adapt to different geometries and configurations, as it needs to fully describe the distribution function $f(\VEC r, \HAT n,X,t)$ of having bacteria at position $\VEC r$, with director $\HAT n$, and normalized concentration of the CheY-P protein $X$ at time $t$. 
A simpler approach is to study the dynamics of slowly-varying fields analog to the Keller--Segel equations \eqref{eqs.ks} for the density field. With that purpose, we will first derive the kinetic equation that describes the dynamics of bacterial suspensions coupled with chemotactic signals with the general model \eqref{langevin_eq_general} and \eqref{tumblingrate}. Then, we will apply the Chapman--Enskog procedure to this kinetic equation,  which is a systematic method to derive the macroscopic equations for the relevant set of slow fields.

\subsection{Kinetic equation}\label{sec.ke}
The distribution function evolves in time by the motion of swimmers, the stochastic evolution of $X$, and  tumbling. All these processes can be captured by the kinetic equation
\begin{align}\label{kinetic_eq}
\frac{\partial f}{\partial t}+V \hat{\mathbf{n}} \cdot \nabla f=\left({\cal F}[f]+\frac{ \Dot{l}}{\tau} \frac{\partial(B f)}{\partial X}\right) + {\cal T}[f],
\end{align}
which describes the temporal evolution of  the distribution function $f(\VEC{r},\hat{\mathbf{n}} ,X, t)$ in  $d$ (2 or 3) spatial dimensions, 
In Eq.~\eqref{kinetic_eq} and subsequent equations, unless necessary for clarity or to avoid ambiguity, we will omit the $X$ and $l$ argument of the functions to simplify the notation
The left hand side (LHS) of the equation describes the persistent motion of bacteria with constant speed $\VEC V=V\HAT{n}$. The parenthesis on the right hand side (RHS) is a Fokker--Planck term corresponding to the Langevin equation~\eqref{langevin_eq_general} that describes the evolution of CheY-P. Here,  
\begin{align}\label{fokker_planck_eq}
{\cal F}[f]\equiv\frac{1}{\tau}\left[\frac{\partial^2 f}{\partial X^2}+\frac{\partial(A f)}{\partial X}\right]
=\frac{1}{\tau}{\cal \widehat F}[f]
\end{align}
is associated to the free evolution of $X$, while the second term in the parenthesis of Eq.~\eqref{kinetic_eq} accounts for the coupling of CheY-P with the ligand. The last term in Eq.~\eqref{kinetic_eq} is 
\begin{align}\label{tumbling_eq}
{\cal T}[f] \equiv\nu_0 C(X)\left[ \int d \hat{\mathbf{n}}^{\prime} w(\hat{\mathbf{n}}^{\prime}\cdot\hat{\mathbf{n}}) f\left(\mathbf{r}, \hat{\mathbf{n}}^{\prime}, X, t\right) -f\right]
\end{align}
and describes the tumbling process as a Lorentz-type equation having a gain term and a loss term, with a tumble rate  determined by
Eq.~\eqref{tumblingrate}. For simplicity, the differential for the integral over the directors will be denoted simply by $d\HAT{n}$ instead of the more formal notation $d^{d-1}\HAT{n}$. The tumbling kernel $w$, which gives the probability for a director  $\hat{\mathbf{n}}^{\prime}$ to change into $ \hat{\mathbf{n}}$, is assumed by isotropy to depend only on the relative angle between the two directors, and it is normalized such that $ \int d \hat{\mathbf{n}}\, w(\hat{\mathbf{n}}^{\prime}\cdot\hat{\mathbf{n}})=1$. In the next sections, we will show that main results will depend only on the first moment of $w$
\begin{align}
\alpha_1 =  \int d \hat{\mathbf{n}}^{\prime} w(\hat{\mathbf{n}}^{\prime}\cdot\hat{\mathbf{n}}) \hat{\mathbf{n}}^{\prime}\cdot\hat{\mathbf{n}},
\end{align}
entering only as a numerical factor in the transport coefficients.
A simple choice for $w$ is to take it  totally isotropic, meaning that the director after tumbling is chosen completely at random in the unit sphere. That is, $w_\text{isotropic}=1/\Omega_d$, where $\Omega_d$ is a the area of the $d$-dimensional unit sphere ($\Omega_2=2\pi$ and $\Omega_3=4\pi$). In this case, $\alpha_{1, \text{isotropic}}=0$. The tumbling of \textit{E.\ coli} is not fully isotropic, with a larger preference toward persisting angles, resulting in $\alpha_{1, \textit{E.coli}}\approx0.33$ \cite{berg1972chemotaxis}. Experiments show that the ligand concentration gradient also affects the tumbling process and the kernel $w$ is also a functional of $l$~\cite{vladimirov2010predicted,saragosti2011directional,saragosti2012modeling}. Here, we will not consider these kind of models, which will be let for future work. 
Kinetic equations similar to Eq.~\eqref{kinetic_eq} have been previously proposed that consider also the internal protein concentration as a relevant variable~\cite{celani2010bacterial,long2017feedback}. In Ref.~\cite{celani2010bacterial} the hydrodynamic equation are derived from the kinetic equation for times much longer than the memory time, resulting in the Keller--Segel equations with values of the transport coefficients that depend on the memory time. In Ref.~\cite{long2017feedback} the kinetic equation is solved for stationary gradients, being able to obtain the chemotactic currents and their transient dynamics.

The kinetic equation can be made dimensionless by taking $\nu_0 = V =1$ and $b\equiv B(0) = 1$, election that fixes the units of time, length, and ligand concentration. In this case, the most important dimensionless parameters of the model are the tumbling rate sensitivity $\lambda\equiv dC(X)/dX|_{X=0}$, the dimensionless memory time $\hat{\tau}=\nu_0\tau$, and the dimensionless intensity of the ligand gradient $\widehat{\nabla l}=bV\nabla l$. The tracking of \textit{E.\ coli} in \cite{figueroa20203d} allowed to fit its stochastic dynamic to the linear models, giving $\hat\tau=4.2$, meaning that bacteria keep the concentration of CheY-P and, therefore, the tumbling rate, constant in average for four tumbling events. As it has not been proven that $\hat\tau$ is larger than one for all flagellated bacteria, we will analyze both the cases of large and small memory times. Finally,  to help readability and the interpretation of the terms, in general we will keep dimensions in what follows.

The eigenvalues of the Fokker--Planck operator, 
\begin{align} \label{eigenf}
{\cal F}[U_n]=-\frac{\gamma_n}{\tau}U_n
\end{align}
satisfy  $0=\gamma_0<\gamma_1<\gamma_2\dots$ and the eigenfunctions are $U_n(X)=u_n(X) \phi(X)$, where 
\begin{align}\label{stat_sol}
\phi(X)  \equiv \phi_0  \frac{e^{\Phi(X)}}{ \Omega_d},
\end{align}
is the stationary solution of Eq.~\eqref{fokker_planck_eq}, with
\begin{align}\label{pot}
\Phi(X) = - \int^X d X'  A(X'),
\end{align}
and $\phi_0$ the normalization constant such that   $u_0=1$ and $\int d\n\int dX \phi(X) u_n(X) u_p(X)=\delta_{np}$~\cite{risken1996}. 
For the linear model $\phi  = \frac{e^{-X^2/2}}{\Omega_d \sqrt{2\pi}}$, $u_n(X)=H_n(X/\sqrt{2})/\sqrt{ 2^n n!}$, with $H_n$  the Hermite polynomial of order $n$ [such that $H_0(x) = 1$, $H_1(x) = 2x$,  . . . ]\cite{arfken2011mathematical}, and $\gamma_n=n$~\cite{villa2023kinetic}.

When $\hat{\tau}\ll 1$, the concentration of CheY-P responds rapidly to variations of the ligand and to fluctuations. Therefore, the relevant macroscopic field is the bacterial density
\begin{align}\label{density_definition}
    \rho(\mathbf{r}, t) \equiv \int d \hat{\mathbf{n}} \int d X f(\mathbf{r}, \hat{\mathbf{n}}, X, t),
\end{align}
which is a conserved slow field. Associated to it, is bacterial current
\begin{align}\label{current_definition}
\mathbf{J}(\mathbf{r}, t) \equiv \int d \hat{\mathbf{n}} \int d X V \hat{\mathbf{n}}  f(\mathbf{r}, \hat{\mathbf{n}}, X, t).
\end{align}
 
On the other side, if the memory time is long, i.e. $\hat\tau \gg 1$, the CheY-P concentration remains constant for several tumbling events.  The ordering of the eigenvalues of ${\mathcal{F}}$ imply that the first CheY-P moment will be the next slowest mode. Therefore, in this case, besides $\rho$ and $\mathbf{J}$, the relevant field for a macroscopic description is the  CheY-P density 
 \begin{align} \label{densityx_definition}
\rho_X(\mathbf{r}, t)\equiv \int d \hat{\mathbf{n}} \int   d X  u_1(X) f(\mathbf{r}, \hat{\mathbf{n}}, X, t),  
\end{align} 
which can be considered as a slowly evolving field. The associated current is
\begin{align} \label{densityx_current_definition}
\mathbf{J}_X(\mathbf{r}, t) \equiv \int    d \hat{\mathbf{n}} \int d X u_1(X) f(\mathbf{r}, \hat{\mathbf{n}}, X, t)  V \hat{\mathbf{n}}.
\end{align}
As in this system there is no spontaneously broken symmetries or critical fields, there are no additional relevant fields~\cite{forster2018hydrodynamic,Chaikin_Lubensky_1995}.

Starting from the kinetic equation \eqref{kinetic_eq}, the objective of this article is to derive macroscopic equations for the relevant fields, as extensions to the Keller--Segel equations. For that, we will apply the Chapman--Enskog method, which is a systematic approach, where the macroscopic dynamical equations are obtained at different time scales \cite{chapman1990mathematical,soto2016kinetic}. In Sect.~\ref{sec.ChE1} we will consider the case of short memory, where $\rho$ is the only conserved field and the application of the Chapman--Enskog method is standard. As it can be anticipated, the Keller--Segel model is recovered  in this regime and the method provides expressions for the transport coefficients from the microscopic dynamics. In Sect.~\ref{sec.ChE2} the case of long memory is worked out. Here, the slow fields are $\rho$ and $\rho_X$ but, as the CheY-P density  is not strictly conserved, we need to apply a modified Chapman--Enskog procedure similar to that used in the study of granular gases, where energy is not conserved \cite{brilliantov2004kinetic,garzo2019granular}. The outcome of this analysis will be a coupled set of equations for $\rho$ and $\rho_X$, where memory manifests in the form of nonlocal responses.

\subsection{General mathematical properties}\label{sec.mathprops}

Before obtaining the conservation equations, we study the properties of the  Fokker--Planck  and tumbling operators of the kinetic equation \eqref{kinetic_eq}. 
First, we define the linear operators  $\mathcal{L}_0$ and $\mathcal{L}$
\begin{align} \label{L_operator}
 \mathcal{L}_0[f] \equiv &\mathcal{F}[f]  +\mathcal{T}[f],\\
 \mathcal{L}[f] \equiv & \mathcal{F}[f] + \frac{ \Dot{l}}{\tau} \frac{\partial (Bf)}{\partial X} +\mathcal{T}[f],
 \end{align}
where in  $\mathcal{L}_0$ we leave apart the coupling term because it has different symmetry properties. Note that the operators $\mathcal{F}$, $\mathcal{T}$, $\mathcal{L}_0$, and $\mathcal{L}$ have all units of inverse of time.

Now we prove that $\mathcal{L}_0$ is hermitian and semidefinite negative
under the scalar product
\begin{align}\label{scalar_prod}
\braket{f}{g} \equiv \int d \n \int d X \phi^{-1}f g,
\end{align}
It is direct to show that 
\begin{multline}\label{fp_property}
\braket{g}{\mathcal{F}[f]} = \frac{1}{\tau}
 \int d \hat{\mathbf{n}}  \int  d X \phi^{-1} g \left[\frac{\partial^2 f}{\partial X^2}+\frac{\partial(A f)}{\partial X}\right]\\
= - \frac{1}{\tau}\int d \hat{\mathbf{n}}  \int  d X \phi^{-1} \left[ \frac{\partial g}{\partial X}  +   Ag\right]
    \left[ \frac{\partial f}{\partial X}  +   Af\right],
\end{multline}
where we consider that the functions must decay in the limits (natural boundary conditions of Ref.~\cite{risken1996}) and we use that $\Phi(X)^\prime = -A(X)$, by using Eq.~\eqref{pot}. For the tumbling part, we get 
\begin{multline} \label{tum_property}
\braket{g}{\tum{}{f}} = - \frac{\nu_0}{2}\int d\HAT{n} \int d \HAT{n}' \int d X  \phi^{-1}C    \\ 
\times w(\hat{\mathbf{n}}^{\prime}\cdot\hat{\mathbf{n}}) \left[ g(\HAT{n}',X)- g(\HAT{n},X) \right]\left[ f(\HAT{n}',X)- f(\HAT{n},X) \right].
\end{multline}
So, assuming that $C(X)>0$, the previous results imply that $\mathcal{F}$ and $\mathcal{T}$ (and hence also $\mathcal{L}_0$) are hermitian and semidefinite-negative.
It is direct to verify by simple substitution that $\mathcal{L}_0[\phi]=0$, implying that $f_{\text{eq} }(\HAT{n},X) = \rho_0 \phi(X)$ is a stationary solution Eq.~\eqref{kinetic_eq} in absence of the ligand. 
 We now show this is the only stationary solution and is associated to the conservation of density.
Let us assume that there is a different  solution of $\mathcal{L}_0[f]=0$, which we write as $f(\HAT{n},X)=\phi(X)\varphi(\HAT{n},X)$.  It is direct that
\begin{align} \label{demphiunique1}
\int d \hat{\mathbf{n}}  \int d X \varphi(\HAT{n},X)\mathcal{L}_0[f] = \braket{f}{\mathcal{F}[f]} + \braket{f}{\mathcal{T}[f]}.
\end{align}
As the LHS vanishes because $f$ belongs to the Ker of $\mathcal{L}_0$ and using that both $\mathcal{F}$ and $\mathcal{L}$ are semidefinite-negative, one obtains that $\braket{f}{\mathcal{F}[f]}=\braket{f}{\mathcal{T}[f]}=0$. As it is standard for the Fokker--Planck operator or from Eq.~\eqref{fp_property}, the only solution of $\braket{f}{\mathcal{F}[f]}=0$ is that $f(\HAT{n},X)=\phi(X) \varphi(\HAT{n})$. Also, from Eq.~\eqref{tum_property}, $\braket{f}{\mathcal{T}[f]}$ can only vanish if $\varphi$ is independent of $\HAT{n}$, that is a constant. With this, we have obtained that there is no other solution apart from $\phi$. This result, together with the semidefinite-negative character of $\mathcal{L}_0$, implies that in absence of a ligand signal, the system relaxes toward the equilibrium distribution $f_{\text{eq} }$ which is isotropic.

The dynamical evolution of the different fields are obtained by computing the moments of the kinetic equation, that is, multiplying Eq.~\eqref{kinetic_eq} by $g(\HAT{n},X)$ and integrating over $\HAT{n}$ and $X$. Conserved fields are those such that the right hand side vanishes for all distribution functions $f$. 
For example, using $g=1$ gives the density field, which is conserved.  
We now show that the bacterial density is the only conserved field for this system. 
For the tumbling operator, we  note that $\int d \hat{\mathbf{n}}\,  d X\, g(\HAT{n},X) \mathcal{T}[f]=\braket{g\phi}{\mathcal{T}[f]}$, which using Eq.~\eqref{tum_property} gives that it vanishes for any $f$ only if $g$ is independent of $\HAT{n}$.  
On the other hand, for the Fokker--Planck term with coupling,
\begin{multline}\label{fp_property2}
	\int d \hat{\mathbf{n}}  \int d X g(\HAT{n},X)\left( \mathcal{F}[f] + \frac{ \Dot{l}}{\tau} \frac{\partial B f}{\partial X}\right) \\
	= - \frac{1}{\tau} \int d \hat{\mathbf{n}}  \int  d X  \frac{\partial g}{\partial X}   
   \left( \frac{\partial f}{\partial X}  + f A    + B \Dot{l} f \right).
\end{multline}
Eq.~\eqref{fp_property2} vanishes for all $f$ only if $g$ is independent of $X$. These two results imply that when computing moments of the kinetic equation \eqref{kinetic_eq}, the right hand side vanishes only if $g$ is simultaneously independent of $X$ and $\HAT{n}$, that is $g=1$. Therefore, density is the only conserved field.

 Importantly,  $\mathcal{T}$ and $\mathcal{F}$ are isotropic (both commute with the rotation operator). This means that, for the angular dependence, solutions of  $\mathcal{L}_0 $  can be expressed as a sum in Fourier series in two dimensions or spherical harmonics in three dimensions. 
But, $\mathcal{T}$ and $\mathcal{F}$ do not commute, and we cannot use a common basis in $X$ for $\mathcal{L}_0$ that simplifies the analysis. 

We choose to use the eigenbasis \eqref{eigenf} , which is a diagonal basis for the Fokker--Planck operator. Then, we will derive the hydrodynamic equations using solutions of the kinetic equation of the form
\begin{align} \label{more_general_solution}
    	 f(\n,X) =    \sum_{n,m }\ket{f_{nm}}  =   \sum_{n,m } f_{nm} e^{in\theta}  U_m(X) .
\end{align}
Although the calculations will be performed both in two and three dimensions, for concreteness, we will present expressions in two dimensions, as in this case. In three dimensions the expansion is analogous, with the use of spherical harmonics instead of the Fourier angular modes. 
 
Applying $\mathcal{L}_0$ over each element of the basis, $\ket{f_{mn}}$, we get,
\begin{multline}
    \label{4.22}
        \mathcal{L}_0\ket{f_{nm}} =  -\frac{\gamma_m }{\tau} U_m(X) e^{i  n \theta}f_{nm} \\+\nu_0 U_m(X) C(X)     \left[\int  d^2 \theta^\prime w(\theta-\theta^\prime)   e^{i  n \theta ' }  -  e^{i  n \theta}\right]f_{nm}.
\end{multline}
Using the inner product defined in Eq.\eqref{scalar_prod}, we find the elements of operator matrix,
\begin{align} \label{4.23}
      F^{n'm'}_{nm} &\equiv{\bra{f_{n'm'}}\mathcal{L}_0\ket{f_{nm}}}  =-\frac{{c_{mm'n}}}{\tau}    \delta_{nn'},
\end{align}
where         
\begin{align}
c_{mm'n}=\nu_0\tau b_{mm'}(1-\alpha_n)+ \gamma_m \delta_{mm'}. \label{eq.defc}
\end{align}
Here
\begin{align}
\alpha_n =  \int d \hat{\mathbf{n}}^{\prime} w(\hat{\mathbf{n}}^{\prime}\cdot\hat{\mathbf{n}}) (\hat{\mathbf{n}}^{\prime}\cdot\hat{\mathbf{n}})^n,
\end{align}
where we note that, by normalization, $\alpha_0=1$.
The matrix 
\begin{align}
\label{matrixbmm_general}
b_{mm'} & = \Omega_d\int_{-\infty}^{\infty}  d X \phi^{-1} C(X) U_m(X) U_{m^\prime}(X),
\end{align}
depends on the explicit chemotactic model and can be easily computed one the functions $A$, $B$, and $C$ are given.  For example, considering the linear model~\eqref{linmodel}, this matrix is
\begin{align}
\label{matrixbmm_linear}
b_{mm'}(\lambda)  =&\frac{1}{\sqrt{2\pi 2^{m+m'} m! m'!}} \nonumber \\
&\times \int_{-\infty}^{\infty} e^{\lambda X-X^2 / 2} H_{m}(X / \sqrt{2}) H_{m'}(X / \sqrt{2}) d X, \nonumber \\
 =&e^{\lambda^2 / 2} \begin{pmatrix}
1 &  \lambda & \cdots \\
 \lambda & \left(1+\lambda^2\right) & \cdots \\
\vdots & \vdots & \ddots
\end{pmatrix},
\end{align}
where we take into account that for linear model, the eigenbasis of the Fokker--Planck equation are the Hermite polinomials. This matrix is nondiagonal, but it becomes close to diagonal for small values of $\lambda$. Then, it is expected that the series \eqref{more_general_solution} can be truncated with few number of Hermite polynomials for small $\lambda$. Consequently, this assumption will be considered to present some explicit expressions. 

In this presentation, we are not considering rotational diffusion as it is a process that does not respond to ligand concentration and is therefore not key to the chemotactic process. It can, however, modify the numerical values, hindering the chemotactic efficiency by introducing additional randomness into the bacterial motion. Anyhow, it can be included by simply adding a term $D_{\text r} \tau n^2$ in Eq.~\eqref{eq.defc}, where $D_{\text r}$ is the rotational diffusion coefficient. For the strain of \textit{E.\ coli} studied in Ref.~\cite{figueroa20203d}, the best fit gave that the dimensionless rotational diffusivity is $\nu_0 D_{\text r}\approx 0.12$, and is therefore subdominant in the analysis.

\section{Chapman--Enskog method for short-time memory} \label{sec.ChE1}
  
\subsection{Chapman--Enskog expansion} \label{c-e_for_rho}
In this section, we consider short memory times, i.e., $\hat\tau \lesssim  1$. This implies that the bacterial density $\rho$ is the only slow field. Integrating Eq.~\eqref{kinetic_eq} 
 over $\HAT{n}$ and $X$, the RHS vanishes and we obtain the conservation equation associated with the density of bacteria,
\begin{align}  \label{dens_cons_eq}
  \pdv{ \rho }{ t} +  \nabla \cdot \VEC{J} = 0 .  
\end{align}
 This equation is not closed for $\rho$, since it depends on the yet unknown $\VEC{J}$ field \eqref{current_definition}, which in turns depends on $f$. The objective here is to  find an explicit functional form  $f=f[\rho]$. To do that, we assume that Eq.~\eqref{kinetic_eq} admits normal solutions, i.e.,  that the normal solutions of the distribution function, $f$, vary over time and space through $\rho$, 
\begin{align}\label{3.10}
   f(\VEC{r},\hat{\mathbf{n}} ,X, t)=  f[ \n , X | \rho(\VEC r,t) ].
\end{align}
To find this functional dependence and, consequently, the macroscopic equation, we assume that spatial gradients are small. For that, we introduce a small bookkeeping   parameter $\varepsilon$, which characterizes the spatial gradients. In this regime, the conserved field is therefore a slowly evolving field. The distribution function can depend on $\rho$ and its gradients, which also vary slowly in space and time.

The distribution function is represented as a series expansion in $\varepsilon$, 
\begin{align}\label{3.11}
    f = f^{(0)} + \varepsilon f^{(1)} +\varepsilon^2 f^{(2)} + \cdots.
\end{align}
This expansion introduces some arbitrariness when fixing the constants and the number of solutions increases. In the Chapman--Enskog method \cite{brilliantov2004kinetic,garzo2019granular}, this arbitrariness is solved by demanding that the $\varepsilon^0 = 1$ order exactly reproduces the hydrodynamic fields, i.e.
\begin{equation}\label{4.4}
    \int d^2 \n  \int d X f^{(k)}  = \rho \delta_{0k}  , \; \forall k.  
\end{equation}

\begin{figure}[htbp]
\centering
\includegraphics[scale=0.38]{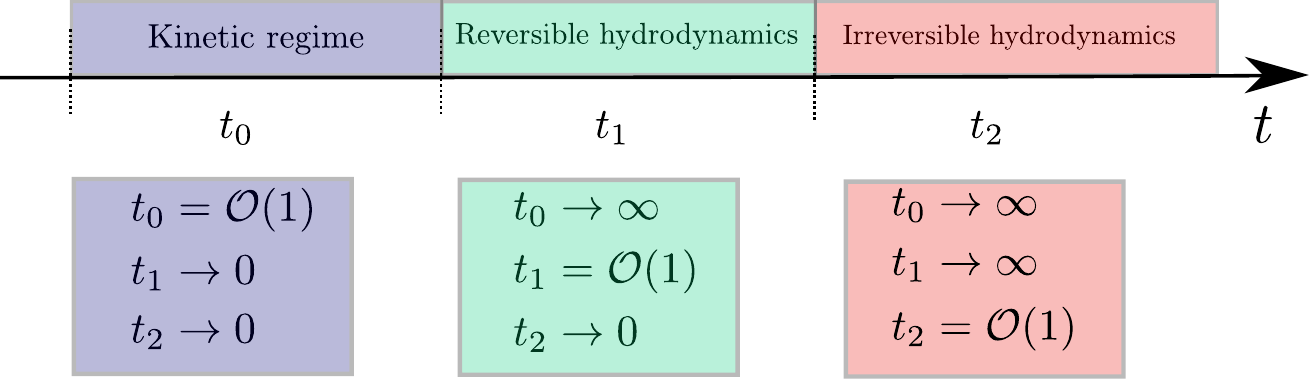}
\caption{Representation of the different time scales of the system. $t_0$ represents the time scale at which kinetic effects are relevant (in our case, tumbling time scale). In this regime, the distribution function evolves towards the local steady-state distribution. $t_1$ corresponds to the reversible hydrodynamic scale. Finally, $t_2$ is the slowest time scale, associated to diffusive processes. }
\label{fig:timescale}
\end{figure}

Next, we introduce a separation of time scales, $t_0=t, \; t_1=\varepsilon t, \; t_2=\varepsilon^2 t, \; \dots$, where we recall that we made time dimensionless by taking $\nu_0=1$. Each temporal variable reflects the dynamics of different time scales (see Fig.~\ref{fig:timescale}). 
For example,  on the tumbling time scale characterized by $t ={\cal O}(1)$,  $t_1$ and $t_2$ will be small and the distribution function will evolve only with $t_0$, with no dependence on $t_1$ or $t_2$. For $t={\cal O}(\varepsilon^{-1})$, on the hydrodynamic time scale linear with gradients, the dynamics will be described only by $t_1$, which takes values of order unity, while $t_0$ has already saturated to large values and $t_2$ still has negligible values. Finally, the slowest scale $t={\cal O}(\varepsilon^{-2})$, describes with $t_2$ the hydrodynamic regime that depends quadratically on the spatial gradients~\cite{soto2016kinetic}.
To change between times scales smoothly and that the expansion is  well-defined mathematically, we impose that the distribution function is regular as $t_i \to 0$ and $t_i \to \infty$, for $i=0,1,2,\dots$.  We can thus write the distribution function as $f= f\left[\n , X | \rho(\VEC{r};t_0,t_1,\dots)\right]$. Using the chain rule for temporal derivatives, we have
\begin{equation}\label{4.3}
\pdv{ f }{t} \rightarrow  \frac{\partial  f }{\partial t_0} +\varepsilon \frac{\partial  f }{\partial t_1}+\varepsilon^2 \frac{\partial  f  }{\partial t_2} +\ldots . 
\end{equation}

Consistent with the previous discussion, we rewrite the kinetic equation \eqref{kinetic_eq} as
\begin{equation} \label{kinetic_hierarquical_equation}
        \pdv{f}{t} + \varepsilon V \n \cdot \nabla{f} = \mathcal{L}_0[f]+ \varepsilon\frac{ \dot{l}}{\tau}\pdv{(Bf)}{X} ,
\end{equation}
where $\varepsilon$ is also introduced in the Lagrangian derivative of $l$, meaning that the driving gradient evolves slowly in space and time, which is a plausible assumption since this field is  externally controlled and molecular diffusion rapidly smoothes any initial large gradient. Substituting  the series expansion \eqref{3.11} and the multiple time scales \eqref{4.3} into  the kinetic equation \eqref{kinetic_hierarquical_equation}, results in a hierarchy of equations  for each order in $\varepsilon$.

\subsection{ {Zeroth order equation}}
 Considering the terms of order zero for $\varepsilon$ in   Eq.~\eqref{kinetic_hierarquical_equation} gives, 
\begin{equation} \label{4.6}
        \pdv{f^{(0)}}{\rho} \pdv{ \rho}{t_0}  =\mathcal{L}_0[f^{(0)}],
\end{equation}
where we used that $f$ depends on time via the density field. Taking the first moment, that is, multiplying Eq.~\eqref{4.6} by 1 and integrating over $\HAT{n}$ and $X$ gives
\begin{align}  \label{4.7}
      \pdv{\rho}{t_0} = 0,
\end{align}
where we use the properties shown in Sec.~\ref{sec.mathprops}.  This result means that $\rho $ does not change in the kinetic time scale, as expected, because neither tumbling or the evolution of $X$ change the particle positions.  Inserting back this result into Eq.~\eqref{4.6}, gives a closed equation for $f^{(0)}$,
\begin{equation} \label{4.8}
       \mathcal{L}_0[f^{(0)}] = 0  .
\end{equation}
In Sec.~\ref{sec.mathprops} we showed that the solution of this equation is $f^{(0)}(\VEC r,\n,t)=\rho(\VEC r,t)\phi(X)$, where $\rho$ has an arbitrary spatio-temporal dependence that will be determined in the next sections when analyzing the kinetic equation to higher orders in $\varepsilon$.

\subsection{{First order equation }}
Following the same steps, we obtain the equation to first order in $\varepsilon$
\begin{equation} \label{hierarchical_Eq_1}
        \pdv{f^{(1)}}{t_0} + \pdv{f^{(0)}}{t_1}+  V \n \cdot \nabla{f^{(0)}}  =   \mathcal{L}_0 \left[f^{(1)}\right] + \frac{\dot{l}}{\tau} \pdv{(Bf^{(0)})}{X}.
\end{equation}
The first term in \eqref{hierarchical_Eq_1} vanishes trivially because $f$ depends on time via $\rho$, and this field does not depend on $t_0$. Taking the first moment of the equation gives
\begin{align}  \label{4.15}
 \pdv{\rho}{t_1} = 0,
\end{align}
where the convective term gives a vanishing contribution by  parity in $\HAT{n}$ and the right hand side vanishes by the properties of $\mathcal{L}_0$ and because $f^{(0)}$ vanishes for $X$ going to infinity.
 This result means that $\rho $ does not evolve neither on this time scale. Considering these results and calculating explicitly the terms that depend on $f^{(0)}$ in Eq.~\eqref{hierarchical_Eq_1}, we find a closed equation for $f^{(1)}$,
\begin{align} \label{4.16}
\mathcal{L}_0[ f^{(1)}] =&   V \n \cdot \left[    \nabla   \rho  + \frac{\rho}{\tau} E_1\nabla l   \right] \phi(X), 
\end{align}
with 
\begin{align}
E_1(X)\equiv \left[A(X)B(X) - \partial_X B(X)\right],
\end{align}
where we used that $\Phi'(X)=-A(X)$.
Here we are considering that the food is stationary, i.e.,  $l \equiv l (\VEC{r})$ to simplify the analysis of the solutions of $f^{(1)}$. Nevertheless, as we show below, the term associated with the explicit time derivative of the ligands does not contribute to the drift-diffusion dynamics, so it can be safely omitted.   Using the isotropy of  $\mathcal{L}_0$, we propose solutions of the form
\begin{align}\label{4.17}
       f^{(1)} = \frac{V}{\nu_0 }  \n \cdot \left[ O(X)    \nabla{\rho}     +  Q(X)  \frac{\rho}{\tau}     \nabla{l} \right],
\end{align}
where the prefactor $\nu_0$ has been included to make the unknowns $O$ and $Q$ dimensionless.
In principle, as in the kinetic theory of gases and granular gases, $O$ and $Q$ are isotropic functions that could depend also on the magnitude of the vector. As $\HAT{n}$ is unitary, this dependence is not present here.

To obtain these functions explicitly, we introduce the expression of $f^{(1)}$ into Eq.~\eqref{4.16}, resulting in the two equations
\begin{subequations}
\begin{align}
\mathcal{ L}_0[O\n] &=  \nu_0 \phi\n, \label{eq.Olineal}\\
\mathcal{ L}_0[Q\n] &=  \nu_0 E_1\phi\n. \label{eq.Qlineal}
\end{align}
\end{subequations}
Notably, the isotropy of the operators imply that for any function $\xi(X)$, $\mathcal{ L}_0[\xi\n]=\frac{\n}{\tau} \mathcal{\widehat L}_0^\prime[\xi]$, where $\mathcal{\widehat L}_0^\prime\equiv  \mathcal{\widehat F} - \nu_0 \tau(1-\alpha_1)C(X)$. Then, expanding in Fokker-Planck eigenfunctions, $O(X)=\sum_m O_m U_m(X)$, and similarly for $Q$, and applying the orthogonality conditions, a set of equations is obtained for the coefficients $O_m$ and $Q_m$, for $m=0,1,\dots$

\begin{subequations}\label{eqsalgOQshort}
\begin{align} 
 & \sum_m c_{mn1}O_m  =  - \nu_0 \tau \delta_{n0},  \label{4.31short}\\
 &    \sum_m c_{mn1} Q_m = -  \nu_0  \tau  \Omega_d \int d X  u_n  E_1 \phi. \label{4.33bshort} 
 \end{align}
\end{subequations}
The solution of these equations depends on the specific model, as specified by the model functions $A$, $B$, and $C$. For example, in the case of the linear model, truncating at $m=1$, that is, keeping two polynomials,
\begin{subequations}
\label{difusion_order_1}
\begin{align} 
            &     O_0 \approx -\frac{1}{e^{\lambda^2/2}}\left[ 1+ \frac{  \lambda^2 \hat\tau e^{\lambda^2/2}}{  (1 +  e^{\lambda^2/2} \hat\tau)} \right] , \\ 
   &     O_1 \approx \frac{  \lambda\hat\tau}{1 +  e^{\lambda^2/2}\hat \tau}, \\       
      &            Q_0 \approx \frac{b\lambda \hat\tau  }{1+e^{\frac{\lambda^2}{2}} \hat\tau} , \\ 
       & Q_1 \approx -\frac{b\hat\tau }{1+e^{\frac{\lambda^2}{2}}\hat \tau} ,
\end{align}
\end{subequations}
where, for simplicity, in the above expressions, we have considered the case of totally isotropic tumbling, that is $\alpha_1=0$.

\subsection{{Hydrodynamic equation}}\label{sec.hydroshortmemory}

The kinetic equation to second order in $\varepsilon$ reads
\begin{equation} \label{eq6}
      \pdv{f^{(2)}}{t_0} +  \pdv{f^{(1)}}{t_1} + \pdv{f^{(0)}}{t_2}+   V \n \cdot \nabla { f^{(1)}} =   \mathcal{L}_0 \left[f^{(2)}\right] + \frac{\dot{l}}{\tau}\pdv{(Bf^{(1)})}{X}    .
\end{equation}
As the density does not depend on $t_0$ and $t_1$, the first and second terms vanish identically. Now, computing the first moment of the equation, gives again a vanishing right hand by the properties of $\mathcal{L}_0$ and because $f^{(1)}$ vanishes for $X$ going to infinity. But, as $f^{(1)}$ is odd in $\HAT{n}$ [Eq.~\eqref{4.17}], the convective term now gives  a finite contribution and the moment equation reads
\begin{align}  \label{eq7}
\pdv{\rho}{t_2}+ \nabla \cdot\VEC{J} = 0,
\end{align}
with the bacterial current $\VEC{J}$  
\begin{align} 
\mathbf{J}(\mathbf{r}, t) &= \int   d \hat{\mathbf{n}} \int d X f^{(1)}(\mathbf{r}, \hat{\mathbf{n}}, X, t) V \hat{\mathbf{n}},\nonumber \label{eq.Jche1}\\  
&=   - D \nabla{\rho} +\mu \rho \nabla{ l},
\end{align}
where  Eq.~\eqref{4.17} for $f^{(1)}$ has been used.
Here,  $D$ is the diffusion coefficient 
\begin{align}
  D = -\frac{V^2\Omega_d}{d\nu_0}\int dX\, O(X)= -  \frac{ V^2O_0 }{d \nu_0}    \label{eq9}
\end{align}
and $\mu$ the mobility,
\begin{align}
    \mu =  \frac{b V^2 \Omega_d}{d \nu_0\tau} \int dX\, Q(X)     
    = \frac{b V^2 Q_0}{d \hat\tau}      \label{eq10}.
\end{align}
The close expressions \eqref{eq9} and \eqref{eq10}, depending on only one coefficient, result from the orthogonality of the base functions. 
Finally, if we had considered an explicit time dependence of the ligand in Eq.~\eqref{4.16}, $f^{(1)}$ would have had an additional term proportional to $\partial_t l$, isotropic in $\HAT{n}$. But this term, would not contribute to the current when integrating over $\HAT{n}$ in \eqref{eq.Jche1}.

In the linear model, $A$ is an odd function and $\mathcal{F}$ is an even operator in $X$. Then, if $C(X)$ had been an even function in $X$ then, by Eq.~\eqref{eq.Qlineal}, $Q$ would have been also even, resulting in a vanishing motility. That is, a finite motility results only due to the odd part of $C$. Indeed, the mechanism of chemotaxis is that $X$ responds to the gradients of ligand becoming positive or negative if moving against or along the gradient, respectively. Then, if the tumbling rate is insensible to this change on sign, the chemotactic mechanism is broken.

\begin{figure}[htb]
\includegraphics[width=\columnwidth]{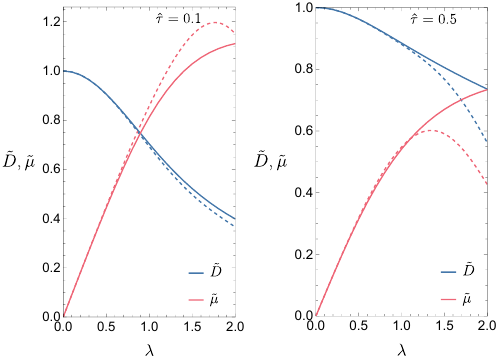}
\caption{Dimensionless diffusion coefficient $\tilde D=d\nu_0D/V^2$ (blue) and chemotactic mobility $\tilde \mu=d\mu/(bV^2)$ (red) for the linear model in the short memory case as a function of the sensitivity $\lambda$ for selected values of the dimensionless memory $\hat\tau$. The solid lines are obtained with the numerical solution of the system of equations \eqref{eqsalgOQshort} truncating it  up to $n=10$ and the dashed lines to the small $\lambda$ approximations using expressions \eqref{difusion_order_1}. In all plots, we consider a fully isotropic tumbling kernel ($\alpha_1=0$).}
\label{fig.D-mu-shortmemory}
\end{figure}

In the case of linear model, explicit expressions can be obtained for the transport coefficients. Figure~\ref{fig.D-mu-shortmemory} presents the $D$ and $\mu$ obtained by solving numerically the system of equations   \eqref{eqsalgOQshort} using the matrix~\eqref{matrixbmm_linear}. A comparison with the small $\lambda$ approximations using expressions \eqref{difusion_order_1} is presented. The approximate expressions have an excellent agreement with the complete solution up to $\lambda\approx 1$, where deviations become to increase importantly, specially for moderate values of the dimensionless memory $\hat\tau$.
Note that the mobility is zero if there are no fluctuations, $\lambda=0$, or there is no coupling with the ligand, $b=0$. Moreover, taking the memoryless limit, $\tau \to 0 $, it results $\mu|_{\tau \to 0} = {bV^2\lambda}/d $,  recovering the mobility expression of the Keller--Segel model \cite{keller1971model}.
The small $\lambda$ approximation for  $\mu$ equals the expression obtained previously from the stationary solution of the kinetic equation for a uniform and constant ligand gradient~\cite{villa2023kinetic}. There, the static mobility was widely analyzed and compared with agent-based simulations obtaining good agreement with the theoretical predictions. 
The diffusion coefficient obtained corresponds to the extension to higher orders in $\lambda$ of the expression obtained in  Ref.~\cite{villa2020run} for the same linear model, but in absence of chemoattractant coupling.  In the small memory limit, $\tau \to 0 $, $D = V^2/(d \nu_0 e^{\lambda^2/2})= V^2/d \langle\nu\rangle$, that is the diffusion coefficient has the form of conventional run-and-tumble model without chemotaxis. In this limit, the concentration of CheY-P becomes a rapidly fluctuating random variable, making bacteria to explore space with an effective tumbling rate equal to its average.


\section{Chapman--Enskog method for long-time memory} \label{sec.ChE2}

\subsection{Equations for the relevant fields}
In this regime of long-time memory, bacteria can travel long distances before adapting to the new environment, for example having a different ligand gradient. This implies that nonlocal effects appear, which can be captured, in principle,  continuing with the Chapman--Enskog procedure to higher orders in the spatial gradients. This is a standard procedure, although cumbersome, that has been applied for example to classical gases \cite{chapman1990mathematical} or rapid granular flows \cite{sela1998hydrodynamic,khalil2014hydrodynamic}, which gives what is called the Burnett and super Burnett equations. Besides some divergencies for the Burnett transport coefficients that appear in classical gases due to mode-coupling long time tails in the correlation functions, the Burnett equations normally have ill-defined boundary conditions that limit their practical applications. Instead of this method, here we proceed to consider that $\rho_X$ as a new relevant field.

As we have shown in the previous sections, there is only one strictly conserved field for the chemotaxis process. But, for the  long memory time regime $\hat\tau\gg 1$, the CheY-P protein density  $\rho_X(\VEC{r},t)$ also evolves slowly and can be considered as a quasi-conserved macroscopic field. This is analogous to the case of rapid granular flows where, for small inelasticities, the kinetic energy is considered also a slowly evolving relevant field and the Chapman--Enskog procedure is applied assuming that the distribution function depends functionally also on the granular temperature~\cite{brilliantov2004kinetic,garzo2019granular}. In rapid granular flows, this results in hydrodynamic-like equations for the density, velocity, and granular temperature, with well-behaved transport laws and excellent agreement with simulations~\cite{brilliantov2004kinetic,garzo2019granular}.

Besides the conservation equation for the density \eqref{dens_cons_eq}, the dynamic equation for $\rho_X$ is obtained by computing the relevant moment of the kinetic equation, that is, multiplying Eq.~\eqref{kinetic_eq} by $u_1$ and integrating over $\HAT n$ and $X$. The results is,
\begin{align}  \label{3.4}
\pdv{\rho_X}{t}+  \nabla \cdot{ \VEC{J}_X}  =  \int d \n \int dX\, u_1(X)  \mathcal{L}[f],
\end{align}
where we used the definition of $\VEC{J}_X$ in Eq.~\eqref{densityx_current_definition}.
To compute  the  RHS of Eq.~\eqref{3.4}, we first note that the contribution of the tumbling operator $\mathcal{T}$ is identically null, because $X$ does not change at the tumbling process.  But, the Fokker--Planck operator gives a finite contribution
\begin{multline}  \label{3.6}
  \int   d \n d X \, u_1(X) \left(\mathcal{F}[f]+\frac{\Dot{l}}{\tau} \frac{\partial (Bf)}{\partial X}\right)  \\
    = -\frac{1}{\tau} \left(\gamma_1 \rho_X +  S_1\pdv{l}{t}  + \VEC{S}_2 \cdot\nabla{l}   \right),
\end{multline}
where
\begin{subequations}\label{3.6b}
\begin{align}  
S_1&=  \int d\n  d X u^\prime_1 B f, \\ 
\VEC{S}_2 &= \int d\n  d X  u^\prime_1 B f V \n. 
\end{align}
\end{subequations}
We have kept the general spatio-temporal dependence of $l$ because, contrary to the case with small memory (Sect.~\ref{sec.ChE1}), here the term $\pdv{l}{t}$  will appear explicitly in the macroscopic equations.

To summarize, the equations for the relevant fields are 
\begin{align}
       \pdv{\rho}{t}+ \nabla \cdot{\VEC{J}} =& 0, \label{rho_eq} \\ 
         \pdv{\rho_X}{t}+ \nabla \cdot{ \VEC{J}_X} =&  -\frac{1}{\tau} \left(\gamma_1 \rho_X +  \pdv{l}{t} S_1 +  \nabla{l} \cdot \VEC{S}_2 \right)
\label{rhox_eq}.
\end{align}
As anticipated, $\rho_X$ is not conserved, but if $\tau$ is large, it evolves slowly on time and becomes a relevant field for the analysis. 
To close these equations, we  proceed as in the previous section and apply the Chapman--Enskog procedure to obtain the transport laws for the fluxes $\VEC{J}$ and $\VEC{J}_X$, and source terms $S_1$ and $\VEC{S}_2$. 

\subsection{Chapman--Enskog method}
To obtain a hierarchy of equations for $\rho$ and $\rho_X$ we first admit the existence of normal solutions. In this regime, this assumption reads
\begin{align}\label{3.10-2}
   f(\VEC{r},\hat{\mathbf{n}} ,X, t)=  f[ \n , X | \rho(\VEC r,t),\rho_X(\VEC r,t) ].
\end{align}
This functional dependence can be locally characterized through gradients of $\rho$ and $\rho_X$, as they also vary slowly. Second, we consider the separation of time scales used in Sec.~\ref{sec.ChE1}, with the introduction of a formal small bookkeeping parameter $\varepsilon$ in front of the spatial gradients and the total temporal derivative of $l$. Three temporal scales will also be used, $t_0=t$, $t_1=\varepsilon t$, and $t_2=\varepsilon^2 t$ (see Fig.~\ref{fig:timescale}). Finally, we expand $f$ in a series in $\varepsilon$ and,  to solve the arbitrariness of the solutions, we impose that the $\varepsilon^0 = 1$ order exactly reproduces both hydrodynamic fields, i.e.
\begin{equation}\label{3.13}
    \int d \n  \,   d X \begin{pmatrix}
        1 \\
u_1  \end{pmatrix}  f^{(k)}  = \delta_{0k} \begin{pmatrix}
        \rho \\
\rho_X \end{pmatrix}, \; \forall k.  
\end{equation}
Expanding $f^{(0)}$ in  $U_n$ and using the orthonormality of these functions, the above condition implies that
\begin{multline} \label{expansionf0}
f^{(0)}(X,\n) = \rho\phi(X) + \rho_X u_1(X)\phi(X) \\+ \sum_{n=0}^1 c_n(\n) u_n(X)\phi(X) + \sum_{n\geq2} d_n(\n) u_n(X)\phi(X),
\end{multline}
with 
\begin{align} \label{eq.conditioncn}
\int c_n(\n) d\n =0.
\end{align}
In the next section, it will be shown that  $c_n$ and $d_n$ vanish for all $n$, but at this stage they are free variables.
 To continue with the Chapman--Enskog procedure, we use Eq.~\eqref{kinetic_hierarquical_equation} and obtain the equations for each order in $\varepsilon$.

\subsection{{Zeroth order equation}}
Considering the terms that are proportional to $\varepsilon^0$, we have        
\begin{equation} \label{3.14}
\pdv{f^{(0)}}{\rho} \pdv{ \rho}{t_0} +\pdv{f^{(0)}}{\rho_X} \pdv{ \rho_X}{t_0} = \mathcal{L}_0[f^{(0)}] .
\end{equation}
Taking the first moment on this equation, we find for the density field
\begin{align}  \label{3.16}
\pdv{\rho}{t_0} = 0,
\end{align}
meaning that, as before, $\rho $ does not  change in the kinetic time scale. Now, taking the moment of $u_1$ in \eqref{3.14}, gives for the the protein density
\begin{align}  \label{3.17}
\pdv{\rho_X}{t_0} =  \int d \n  \int d X \, u_1 \mathcal{L}_0[f^{(0)}] .
\end{align}
Let's work in detail the different terms of the RHS.
For that, we first note that for any $f^{(0)}$, the integral of the tumbling term vanishes. Now, for the Fokker--Planck term, we have that
\begin{align}
{\cal F}[f^{(0)}] =& {\cal F}\bigg[ \rho\phi(X) + \rho_X u_1(X)\phi(X) \nonumber\\
&+ \sum_{n=0}^1 c_n(\n) u_n(X)\phi(X) + \sum_{n\geq2} d_n(\n) u_n(X)\phi(X)
\bigg] \nonumber\\
=&-\frac{\gamma_1}{\tau}\rho_X u_1(X)\phi(X) -\frac{\gamma_1}{\tau} c_1(\n) u_1(X) \phi(X) \nonumber \\
&-  \sum_{n\geq2} \frac{\gamma_n}{\tau} d_n(\n) u_n(X)\phi(X),
\end{align}
where we used that ${\cal F}[\phi]=0$ and ${\cal F}[u_n\phi]=-\frac{\gamma_n}{\tau}u_n\phi$. 
Then,
\begin{align}
\int d \n  \int   d X u_1 {\cal F}[f^{(0)}] = -\frac{\gamma_1}{\tau} \rho_X,
\end{align}
where we used the orthonormality of $u_n$ and Eq.~\eqref{eq.conditioncn}.
With all these terms, we obtain
\begin{align} \label{eq.rhoxt0}
\pdv{\rho_X}{t_0} =  -\frac{\gamma_1}{\tau} \rho_X.
\end{align}
This implies that $\rho_X$ is not conserved but yet it evolves slowly because $t_0/\tau \ll1$. 
 Substituting these results in \eqref{3.14} gives an equation for $f^{(0)}$ 
\begin{equation}\label{3.18}
        -\frac{\gamma_1}{\tau} \pdv{f^{(0)}}{\rho_X} \rho_X = \mathcal{L}_0[f^{(0)}].
\end{equation} 
This equation is linear and isotropic. As there is no preferred direction, $f^{(0)}$ should be isotropic as well. This implies that in Eq.~\eqref{expansionf0}, $c_n$ and $d_n$ should not depend on the angle. The condition Eq.~\eqref{eq.conditioncn} implies then that $c_0=c_1=0$.  
Replacing  the resulting expansion in Eq.\ \eqref{3.18} gives
\begin{align}
- \sum_{n\geq2} d_n \left(\frac{\gamma_n}{\tau} +\nu_0  C\right)u_n\phi=0.
\end{align}
As $\gamma_n>0$ and $\nu_0 C(X)\geq0$, we obtain finally that $d_n=0$ for all $n$. With this, the solution is
\begin{align} \label{f0}
f^{(0)}(\VEC r,\n,t)=\rho(\VEC r,t)\phi(X) + \rho_X(\VEC r,t) u_1(X)\phi(X).  
\end{align}

\subsection{ {First order solutions}}
To first order in $\varepsilon$, we obtain the same equation as in the previous section [Eq.~\eqref{hierarchical_Eq_1}]. 
 Taking the moments of $\rho$ and $\rho_X$ it is found that
\begin{align}  
\pdv{\rho}{t_1} &= 0, \label{eq.rhot1}\\
\pdv{\rho_X}{t_1} &=  -\frac{1}{\tau} \left( S_1^{(0)} \pdv{l}{t}  +  \VEC{S}_2^{(0)} \cdot \nabla{l}  \right),\label{eq.rhoxt1}
\end{align}
where, $S_1^{(0)}$ and $\VEC{S}_2^{(0)}$ are computed using Eqs.~\eqref{3.6b} with $f^{(0)}$. By the isotropy of $f^{(0)}$, it is direct that $\VEC{S}_2^{(0)}=0$, while $S_1^{(0)}=g_1\rho + g_2\rho_X$, where
\begin{subequations}
\begin{align}
g_1 &=\int d\n \int dX u_1'  B\phi ,\\ 
g_2 &=\int d\n \int dX u_1'  u_1 B\phi  
\end{align}
\end{subequations}
are numerical coefficients that depend on the specific model via $A$ and $B$, and $u_1'= \pdv{u_1}{X}$. For the linear model, $g_1=b$ and $g_2=0$.

Using the slaving of the distribution functions to the fields and Eqs.~\eqref{3.16}, \eqref{eq.rhoxt0}, \eqref{eq.rhot1}, and \eqref{eq.rhoxt1} to express the temporal derivatives, Eq.~\eqref{hierarchical_Eq_1} reads
\begin{multline}\label{3.29pre}
{ \left(   \frac{\gamma_1}{\tau}\rho_X \pdv{}{\rho_X} +   \mathcal{L}_0 \right)} f^{(1)} 
  =V \n \cdot \nabla f^{(0)}        \\      - \frac{1}{\tau}\pdv{l}{t}  (g_1\rho + g_2\rho_X) u_1\phi- \frac{\dot{l}}{\tau}\pdv{B f^{(0)}} {X} .
\end{multline}
Introducing the explicit form of $f^{(0)}$ [Eq.~\eqref{f0}], we obtain that $f^{(1)}$ satisfies the equation, 
\begin{multline}\label{3.29}
{ \left(   \frac{\gamma_1}{\tau}\rho_X \pdv{}{\rho_X} +   \mathcal{L}_0 \right)} f^{(1)} 
  =              
  \frac{1}{ \tau}  \left(\rho  E_2 +  \rho_X  E_4  \right)   \phi \pdv{l}{t} \\
  + V \n \cdot \bigg[    \nabla   \rho  +
  u_1  \nabla \rho_{X}     
  +  \frac{1}{\tau}\left(\rho E_1+  \rho_X E_3 \right) \nabla l   \bigg] \phi,
\end{multline} 
with 
\begin{subequations}\label{eqs.definicionesaux}
    \begin{align} 
E_1(X) &\equiv     A(X)B(X) -B^\prime(X), \\
E_2(X) &\equiv  E_1(X)- g_1 u_1(X)      , \\
E_3(X) &\equiv  u_1(X)E_1(X)  - u^\prime_1(X) B(X), \\
E_4(X)  &\equiv E_3(X)  - g_2 u_1(X)  .
 \label{2}
    \end{align}
\end{subequations}
For the linear case,  $E_1(X)=bX$, $E_2(X)=0$ and $E_3(X)=E_4(X)= b(X^2-1)$. 

To solve Eq.~\eqref{3.29}, we use that $f^{(1)}$ must be linear on $\rho$ and $\rho_X$.  Since the operator $\frac{\gamma_1}{\tau}\rho_X\pdv{}{\rho_X}+\mathcal{L}_0$ is isotropic, the solutions are of the form
\begin{multline}\label{3.30}
 f^{(1)}\left[\n,X|\rho,\rho_X\right]  =    \frac{  (\rho M +\rho_X N )}{  \nu_0\tau}  \pdv{l}{t}  \\
 +\frac{    V}{  \nu_0} \n \cdot\bigg[O   \nabla{ \rho}  + P    \nabla{ \rho_X}  
 +\frac{1}{\tau}\left(\rho Q + \rho_X R\right)    \nabla{l} \bigg],
\end{multline}
where $M$, $N$, $O$, $P$, $Q$ and $R$ are yet unknown functions of $X$, and the prefactors $\nu_0$ where introduced to make these functions dimensionless. To obtain them, we replace this expression for $f^{(1)}$ into Eq.~\eqref{3.29}, resulting in
\begin{subequations}\label{eqsdiffM-Q}
    \begin{align} 
     \mathcal{\widehat F}[M] = & \nu_0 \tau   E_2 \phi , \label{3.33d} \\
\gamma_1N +     \mathcal{\widehat F}[N] = & \nu_0 \tau  E_4  \phi, \label{3.33p} \\       
\mathcal{\widehat L}^\prime_0 [O] = &  \nu_0 \tau  \phi,  \label{3.31}\\
  \mathcal{\widehat L}^\prime_0[P] = & \nu_0 \tau   u_1  \phi, \label{c} \\
  \mathcal{\widehat L}^\prime_0 [Q] = & \nu_0\tau   E_1  \phi, \label{3.33b} \\
\gamma_1 R +   \mathcal{\widehat L}^\prime_0 [R] = & \nu_0\tau   E_3  \phi \label{eq.paraR}, 
    \end{align}
\end{subequations}
with $\mathcal{\widehat L}_0^\prime\equiv  \mathcal{\widehat F} - \nu_0 \tau(1-\alpha_1)C(X)$. 
For Eqs.~\eqref{eqsdiffM-Q} we propose a series expansions in Fokker-Planck eigenfunctions, $M(X)=\sum_m M_m U_m(X)$, and similarly for $N,O,P,Q$, and $R$. So applying the orthogonality condition, a set of equations are obtained for the expansion coefficients, 
\begin{subequations}\label{eqsalgM-Q}
    \begin{align} 
  &  M_n =  -\frac{\nu_0  \tau \Omega_d}{\gamma_n} \int d X  u_n E_2 \phi , \label{4.33d} \\
  & N_n =  -\frac{\nu_0  \tau \Omega_d}{(\gamma_n-\gamma_1)} \int d X  u_n E_4 \phi, \label{4.33p} \\       
 & \sum_m c_{mn1}O_m  =  - \nu_0 \tau  \delta_{n,0},  \label{4.31}\\
 &  \sum_m c_{mn1}P_m  =  - \nu_0 \tau  \delta_{n,1}, \label{4.33a} \\
 &    \sum_m c_{mn1} Q_m = -  \nu_0  \tau  \Omega_d \int d X  u_n  E_1  \phi, \label{4.33b} \\
 &  \sum_m c_{mn1}  R_m - \gamma_1 R_n=  -\nu_0  \tau  \Omega_d \int d X  u_n  E_3 \phi. 
    \end{align}
\end{subequations}
Once the model functions $A$, $B$, and $C$ are provided, these equations can be directly solved to obtain $f^{(1)}$. For example, in Appendix~\ref{app.solutionlinear} we provide the solutions for the linear model.

\subsection{ {Hydrodynamics equations}}
 
Having determined  $f^{(1)}$, the  hydrodynamic-like equations for $\rho$ and $\rho_X$ are obtained taking the associate moments of the kinetic  equation to order $\varepsilon^2$,
\begin{equation} \label{eq6-ChE2}
      \pdv{f^{(2)}}{t_0} +  \pdv{f^{(1)}}{t_1} + \pdv{f^{(0)}}{t_2}+   V \n \cdot \nabla   f^{(1)}  =   \mathcal{L}_0 \left[f^{(2)}\right] + \frac{b\dot{l}}{\tau}\pdv{f^{(1)}}{X}    .
\end{equation}
From the first moment we obtain the equation for the bacterial density
\begin{align}  \label{eq7-ChE2}
   \pdv{\rho}{t_2}+ \nabla \cdot \VEC{J}  = 0,
\end{align}
with the  bacterial current $\VEC{J}$ given by
\begin{align} 
\mathbf{J} (\mathbf{r}, t) &= \int   d \hat{\mathbf{n}} \int d X f^{(1)}(\mathbf{r}, \hat{\mathbf{n}}, X, t) V \hat{\mathbf{n}}, \nonumber\\  
&=    - D_{11} \nabla \rho + D_{12} \nabla\rho_X +(\mu_{11} \rho -\mu_{12} \rho_X) \nabla l   ,\label{eq8}
\end{align}
where we used the expression \eqref{3.30} for $f^{(1)}$.
Thanks to the orthogonality of $U_m$, the transport coefficients $D_{11}$, $D_{12}$, $\mu_{11}$, and $\mu_{12}$ are given in terms of single coefficients of the expansions
\begin{subequations}\label{eqs.coefstr_1}
\begin{align}
D_{11} &=  -\frac{V^2 O_0 }{d \nu_0}  ,\label{eq9.1A}\\
D_{12} &=\frac{V^2 P_0}{d\nu_0},\label{eq9.2A}\\
 \mu_{11} &=\frac{V^2 Q_0}{d\hat\tau} ,\label{eq9.3A}\\
 \mu_{12} &=-\frac{V^2 R_0}{d\hat\tau }.\label{eq9.4A}
\end{align} 
\end{subequations}
Note that $D_{12}$, $\mu_{11}$, and $\mu_{12}$  depend in principle on $l$, because $A(X,l)$, $B(X,l)$ and $C(X,l)$ could be  functions of the ligand. 

Computing the moment of $u_1$ of Eq.~\eqref{eq6-ChE2}, we obtain the macroscopic equation for $\rho_X$,
\begin{align}  \label{eq13}
   \pdv{\rho_X}{t_2}+ \nabla \cdot{ \VEC{J}_X} =  -\frac{1}{\tau} \left( \pdv{l}{t} S_1^{(1)} +  \nabla{l} \cdot \VEC{S}_2^{(1)} \right),
\end{align}
where the protein current $\VEC{J}_X$ is
\begin{align} 
\mathbf{J}_X (\mathbf{r}, t) &= \int   d \hat{\mathbf{n}} \int d X f^{(1)}(\mathbf{r}, \hat{\mathbf{n}}, X, t) u_1(X) V \hat{\mathbf{n}}, \nonumber\\
&=   D_{21} \nabla{\rho} - D_{22} \nabla{\rho_X} -(\mu_{21} \rho - \mu_{22}\rho_X ) \nabla{ l}.\label{eq14}
\end{align}
The transport coefficients $D_{21}$, $D_{22}$, $\mu_{21}$, and $\mu_{22}$  are also given in terms of single terms of the expansion,
\begin{subequations}\label{eqs.coefstr_2}
\begin{align}
D_{21} &=  \frac{V^2 O_1 }{d \nu_0}  ,\label{eq9.1B}\\
D_{22} &=-\frac{V^2 P_1 }{d\nu_0},\label{eq9.2B}\\
 \mu_{21} &=-\frac{V^2 Q_1}{d\hat\tau} ,\label{eq9.3B}\\
 \mu_{22} &=\frac{V^2 R_1}{d\hat\tau }.\label{eq9.4B}
\end{align} 
\end{subequations}
Finally, the source terms are computed using $f^{(1)}$ and given by
\begin{align}
S_1^{(1)} &= (g_3 \rho +g_4\rho_X)\pdv{l}{t},  \\
\VEC{S}^{(1)}_2 &= g_5 \nabla \rho + g_6\nabla\rho_X + (g_7\rho+g_8\rho_X)\nabla l,
\end{align}
with
\begin{subequations}\label{eqsg38}
\begin{align}
g_3&=  \frac{1}{ \nu_0\tau} \int dX u_1^\prime B M, \\
g_4&=  \frac{1}{ \nu_0\tau} \int dX  u_1^\prime B N, \\ 
g_5 &=  \frac{V^2}{ d\nu_0} \int dX  u_1^\prime B O, \\ 
g_6 &=  \frac{V^2}{ d \nu_0} \int dX   u_1^\prime B P, \\
g_7 &=  \frac{V^2}{ d \nu_0\tau} \int dX  u_1^\prime B Q, \\
g_8 &=  \frac{V^2}{ d \nu_0\tau} \int dX  u_1^\prime B R, 
\end{align}
\end{subequations}
which are numerical values that depend on the model for $A$, $B$ and $C$. 
In the linear model, $B(X)=b$ and $u_1'(X)=H_0(X/\sqrt{2})$, resulting in $g_3=g_4=0$, $g_5=bV^2 O_0/(d\nu_0)=-bD_{11}$, $g_6=bV^2 P_0/(d\nu_0)=bD_{12}$, $g_7=bV^2 Q_0/(d\hat\tau)=b\mu_{11}$, and $g_8=bV^2 R_0/(d\hat\tau)=-b\mu_{12}$.

The election of the signs in Eqs.~\eqref{eq8} and \eqref{eq14} is such that the transport coefficients are all positive for the lineal model. In the general case, $D_{11}$ and $D_{22}$ are proportional to  $-[c^{-1}]_{00}$ and $-[c^{-1}]_{11}$, respectively, which are the matrix elements of $\mathcal{L}_0^{-1}$. As $\mathcal{L}_0$ is semidefinite negative, it results therefore that $D_{11}$ and $D_{22}$ are always positive, regardless of the model.
Eqs.~\eqref{4.31} and \eqref{4.33a} give that $O_1=\nu_0\tau\Omega_d[c^{-1}]_{10}$ and $P_0=\nu_0\tau\Omega_d[c^{-1}]_{01}$. As the $c$ matrix is symmetric, it turns out that $D_{12}=D_{21}$, which is a sort of  \textit{Onsager reciprocal relation}  \cite{de1984non} for this nonequilibrium system. 
In the linear model, another symmetry emerges. Indeed, in this case the mobility and diffusion coefficients are related by $\mu_{11}=b D_{12}/\tau$ and $\mu_{21}=b D_{22}/\tau$ [Eqs.~\eqref{eq9.2A}, \eqref{eq9.3A}, \eqref{eq9.2B}, and \eqref{eq9.3B}], and therefore $ \mu_{21}=\mu_{11} D_{22}/D_{12}$.

To complete the Chapman--Enskog method, the general macroscopic equation for $ \rho$ and $\rho_X$ are obtained by summing the dynamics for all orders in $\varepsilon$  and replacing $t_r= \varepsilon^r t$, resulting in
\begin{align}
\pdv{\rho}{t} + \nabla \cdot\VEC{J} &= 0 ,\label{complet_rho_eq}\\
\pdv{\rho_X}{t}+ \nabla \cdot{ \VEC{J}_X} =&  -\frac{1}{\tau} \left(\gamma_1 \rho_X +  \pdv{l}{t} S_1 +  \nabla{l} \cdot \VEC{S}_2 \right), \label{complet_rhox_eq}
\end{align}
where, using the relations between the transport coefficients, the currents and sources are written as
\begin{subequations}\label{eqs.constlaws}
\begin{align} 
\mathbf{J} &=    - D_{11} \nabla \rho + D_{12} \nabla\rho_X +(\mu_{11} \rho -\mu_{12} \rho_X) \nabla l,  \label{eqJJxfinal.J}\\
 \mathbf{J}_X &=   D_{12} \nabla{\rho} - D_{22} \nabla{\rho_X} -(\mu_{21} \rho - \mu_{22} \rho_X ) \nabla{ l}, \\
 S_1 &= g_1\rho + g_2\rho_X + (g_3 \rho +g_4\rho_X)\pdv{l}{t},  \\
\VEC{S}_2 &= g_5 \nabla \rho + g_6\nabla\rho_X + (g_7\rho+g_8\rho_X)\nabla l.
\end{align}
\end{subequations}
The dynamical equations \eqref{complet_rho_eq} and \eqref{complet_rhox_eq} with the constitutive relations  Eqs.~\eqref{eqs.constlaws} consitutue the main result of this article. They describe the dynamical evolution of bacterial suspensions coupled with chemotactic signals when the memory time of the chemotactic circuit is appreciable. The equations keep their form for any microscopic model for the bacterial response as long as it is characterized by a single long time scale $\tau$. That is, these equations are valid for times longer than the other, smaller, internal time scales.
Finally, in the linear model, the substitution of the  constants $g_i$ give $S_1=b\rho$ and $\VEC S_2=b\VEC J$, resulting in a simplification of the equations.

\begin{figure*}[htb]
\includegraphics[width=2\columnwidth]{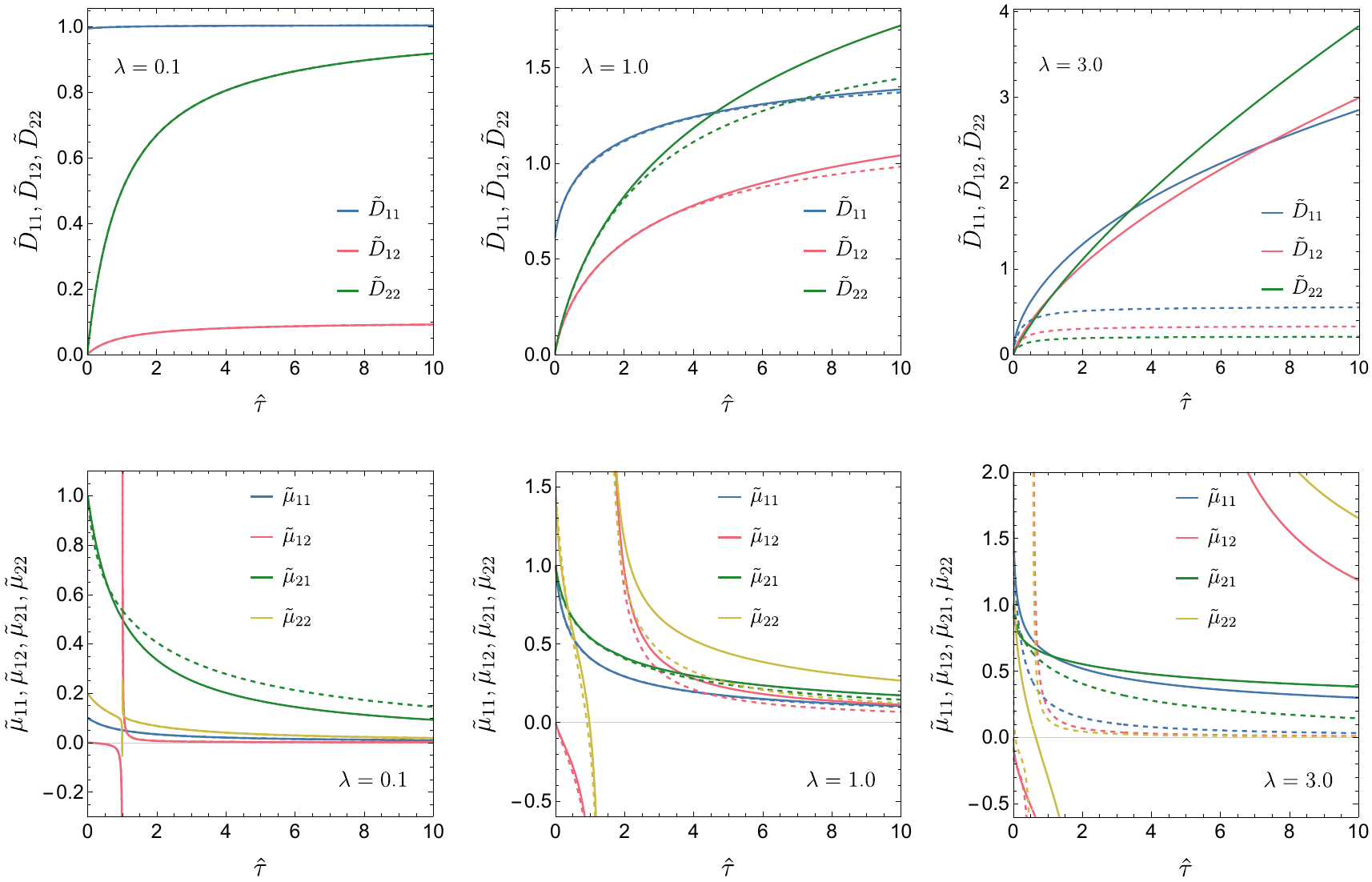}
\caption{Transport coefficients for the linear model in the long memory case as a function of the dimensionless memory $\hat{\tau}$ for selected values of the sensitivity $\lambda$. Top: dimensionless diffusion coefficients $\tilde D_{11}=d\nu_0D_{11}/V^2$ (blue), $ \tilde D_{12}=d\nu_0D_{12}/V^2$ (red), and $\tilde{ D}_{22}=d\nu_0D_{22}/V^2$ (green). 
Bottom: dimensionless chemotactic mobilities  $\tilde \mu_{11}=d\mu_{11}/(bV^2)$ (blue), $\tilde \mu_{12}=d\mu_{12}/(bV^2)$ (red), $\tilde {\mu}_{21}=d{\mu}_{21}/(bV^2)$ (green), and $\tilde {\mu}_{22}=d{\mu}_{22}/(bV^2)$ (yellow). 
The solid lines correspond to evaluating the transport coefficients using the numerical solution of the system of equations \eqref{eqsalgM-Q} truncating it  up to $n=10$ and the dashed lines to the small $\lambda$ approximations using the expressions \eqref{eqM.orden2}, \eqref{eqN.orden2}, and \eqref{eqP.orden2}. In all plots, we consider a fully isotropic tumbling kernel, with $\alpha_1=0$.}
\label{fig.transportcoeff.ChE2}
\end{figure*}

Figure \ref{fig.transportcoeff.ChE2} present the transport coefficients for the linear case with a fully isotropic tumbling kernel ($\alpha_1=0$), obtained by solving numerically the linear equations \eqref{eqsalgM-Q}, and comparing them with the analytic expressions for small $\lambda$ given in App.~\ref{app.solutionlinear}.
It should be remarked that $\mu_{12}$ and $\mu_{22}$ diverge for specific values of $\hat\tau$ that depend on $\lambda$. This divergence, which happens for all values of $\alpha_1$, cannot be physical and can be attributed to the limitations of the linear chemotactic model or to a lack of full separation of the kinetic modes, a necessary condition for the Chapman--Enskog method, as observed in some models of granular gases~\cite{santos2003transport,brey2010breakdown}. As the kinetic theory does not present such divergence, a possible workaround would be to consider other fields in the description, associated to higher modes $u_2$, $u_3$, \dots.

\section{Analysis of results}\label{sec.analysis}

\subsection{Relation with the short-memory case}\label{sec.longvsshort}
Although the hydrodynamic equations for $\rho$ and $\rho_X$ were obtained assuming that $\hat\tau$ is large, it is possible to analyze them even in the short memory limit and compare the results with those obtained in Sec.~\ref{sec.hydroshortmemory}. 
The analytic structure of Eqs.~\eqref{eqsdiffM-Q} give that in the limit $\tau\to0$, $P$ and $R$ vanish, while $O$ and $Q$ remain of order 1 (see Appendix~\ref{app.limits}). As a result, in this limit $D_{12}$ and $\mu_{12}$ vanish, while $D_{11}$ and $\mu_{11}$ are finite, implying that $\rho$ decouples from $\rho_X$ and Eqs.~\eqref{eq7} and \eqref{eq.Jche1} of the short memory case are recovered. Also, as the equations for $D_{11}$ and $\mu_{11}$ are the same as for $D$ and $\mu$ of the short memory case, respectively, the transport coefficients are the same: $D=D_{11}$ and $\mu=\mu_{11}$. 
Also, in this limit, $\rho_X$ relaxes rapidly such as to vanish the right hand side of Eq.~\eqref{complet_rhox_eq}.

\subsection{Homogeneous stationary solution}

The kinetic equation is linear in the distribution function, but the response to the chemoattractant is generally nonlinear, as was previously discussed in the analysis of the kinetic equation for the linear model in Ref.~\cite{villa2023kinetic}. Indeed, the response cannot be linear to infinitely large gradients as the current saturates to the maximum possible value $|\VEC J|=\rho V$, when all swimmers point in the same direction. At the level of the hydrodynamic equations derived here, the expansion in $\varepsilon$   corresponds also to an expansion in $\nabla l$ and the fluxes have been obtained up to linear order in the gradients. Nonlinear contributions could be obtained going to the Burnett and super-Burnett orders in the Chapman--Enskog expansion \cite{chapman1990mathematical}.

To study the system response, we first focus on the simplest case where the ligand gradient is homogeneous and stationary $|\nabla l   |\equiv l_0$, and we look for the steady state. Equations \eqref{complet_rho_eq} and \eqref{complet_rhox_eq} give
\begin{align}
|\VEC{J}| =& \mu_{11} \rho_0  l_0 + \frac{\mu_{12} g_7  \rho_0 l_0^2 }{\gamma_1+g_8l_0^2}, \label{eq:current_start_hom}\\
\rho_{X} =&  -\frac{g_7  \rho_0 l_0^2 }{\gamma_1+g_8l_0^2}. \label{eq:densityx_stat_hom}
\end{align}
Consistent with the above discussion that nonlinear contribution to the fluxes appear on higher orders of the Chapman--Enskog method, the second term in Eq.~\eqref{eq:current_start_hom} should be discarded. That is, the bacterial flux is proportional to the ligand gradient, with a mobility $\mu_{11}$ equal to the short memory case. This is a consequence of studying the stationary regime, where, although slow, $\rho_X$ has had the time to relax. In the  steady state, to dominant order, the mean value of the CheY-P protein is $\langle X \rangle\equiv \rho_X/\rho = -g_7 l_0^2$. As in the linear model $g_7>0$, this results shows  that on average bacteria move in the same direction of ligand gradient.

Still in the case of stationary ligand profiles, the nonlinear coupling between $\rho$ and $\rho_X$ with $\nabla l$ in the fluxes, together with the memory effects, can generate interesting nonlinear responses. According to Eq.~\eqref{langevin_eq_general}, if bacteria are subject to a large ligand gradient beyond the linear regime, $\langle X \rangle$ will take large negative values, which  cannot be obtained by Eqs.~\eqref{complet_rho_eq} and \eqref{complet_rhox_eq}. Nevertheless, if after that, bacteria enter into a region with moderate gradients, where this theory is valid, according to Eq.~\eqref{eqJJxfinal.J}, the bacterial current will be given by $\VEC J = \mu_\text{eff} \nabla l$, where the effective mobility is
\begin{align} \label{3.72}
    \mu_\text{eff} =  \mu_{11} -\mu_{12} \langle X\rangle.
\end{align}
which is larger than the stationary mobility $\mu_{11}$. This mobility is transitory, on the time scale $\tau$ needed for $\rho_X$ to relax to the nonlinear steady state value \eqref{eq:densityx_stat_hom}. We recall that as the unnormalized CheY-P concentration is bound to limits, $X$ and therefore the effective mobility will  be limited as well. 

\subsection{Linear spatio-temporal response}
For the case of a forcing with spatio-temporal structure, the general nonlinear response is complex and normally can only be obtained by solving numerically the hydrodynamic equations. As a relevant case, we study the linear response of the system to small perturbations of the chemoattractant. For that we consider it to be of the form $l(\VEC{r},t)= l_0 + \eta l_1 e^{i(\VEC{k}\cdot\VEC{r}- \omega t )}$, where $\eta$ is a small parameter, and study the linear response of $\rho$ and $\rho_X$, in which case both fields are also  described by Fourier modes
\begin{subequations}
\begin{align}
    \rho =& \rho_0 + \eta\rho_1 e^{i(\VEC{k}\cdot\VEC{r}-
\omega t )}, \label{eq:ro_perturbed} \\ 
    \rho_X =& \rho_{X0} + \eta \rho_{X1}e^{i(\VEC{k}\cdot\VEC{r}-
\omega t )}.  \label{eq:rox_perturbed}
\end{align} 
\end{subequations}
To zeroth order in $\eta$, Eqs.~\eqref{complet_rho_eq} and \eqref{complet_rhox_eq} give that $\rho_{X0}=0$ and $\rho_0$ is arbitrary, fixed by the conserved mass initial condition. Expanding the equations to first order in $\eta$, it is found that $\rho_1=\rho_0\Psi_\rho(k,\omega)  l_1$ and $\rho_{X1}=\rho_0{\Psi}_X(k,\omega)  l_1$, where the response functions are directly obtained from the linearized hydrodynamics and are given in Appendix~\ref{expressions.gral}.
To gain insight, we consider a stationary perturbation, in which case the response functions must be evaluated at $\omega=0$ and the static response functions simplify to
\begin{subequations}
\begin{align}\label{eq.responselonretz}
\Psi_\rho(k,0)&=  \frac{\psi_0}{ \left[1+\left({k}/{k_0}\right)^2\right]}+\psi_1,\\
\Psi_X(k,0)&= -\frac{(D_{11}\psi_0/D_{12}) (k/k_0)^2}{ \left[1+\left({k}/{k_0}\right)^2\right]},
\end{align}    
\end{subequations}
with 
\begin{subequations}
\begin{align}
\psi_0&=\frac{D_{12}(D_{11}\mu_{21}-D_{12}\mu_{11})}{D_{11}(D_{11}D_{22}-D_{12}^2)},\\
\psi_1&=\frac{D_{22}\mu_{11}-D_{12}\mu_{21}}{(D_{11}D_{22}-D_{12}^2)},\\
 k_0 &=  \sqrt{\frac{D_{11}\gamma_1}{\tau \left(D_{11}D_{22}-D_{12}^2\right)}}.
\end{align}
\end{subequations}
The response function $\Psi(k,0)$, has a constant term, associated to a local response, added to a Lorentzian form, indicating a nonlocal response with a smoothing length $L_0=k_0^{-1}$. 
In the memoryless limit ($\tau\to 0 $) the density response function becomes $\Psi(k,0) = \psi_0+\psi_1= \mu_{11}/D_{11}=\mu/D$, independent of $k$, indicating that the response is completely local and we recover the response of the Keller--Segel model~ \cite{keller1970}. 

\begin{figure}[t!]
\begin{center}
\includegraphics[width=\columnwidth]{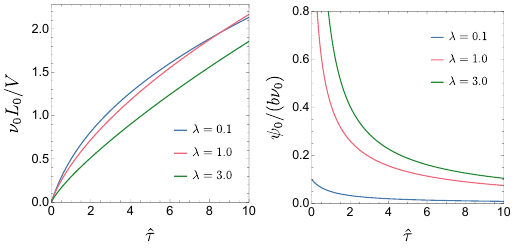}
\end{center}
\caption{Smoothing length $L_0$  (left) and  amplitude $\psi_0$ (right)  of the static response function [Eq.~\eqref{eq.responselonretz}] for the linear model as a function of the dimensionless memory time $\hat{\tau}$ for $\lambda=0.1$ (blue), 1.0 (red), and 3.0 (green). The solid lines correspond to evaluating the transport coefficients using the numerical solution of the system of equations \eqref{eqsalgM-Q} truncating it  up to $n=10$. In all plots, we consider a fully isotropic tumbling kernel, with $\alpha_1=0$.}
\label{fig.k0psi0}
\end{figure}

In the linear model, the symmetry $\mu_{21}=\mu_{11} D_{22}/D_{12}$ of the transport coefficients result in that $\psi_0=\mu_{11}/D_{11}$ and $\psi_1=0$, implying that the density response is purely nonlocal. 
Figure \ref{fig.k0psi0} shows the amplitude of the density response $\psi_0$ and the smoothing length $L_0$ for the linear model using the numerical solution of the equations \eqref{eqsalgM-Q} for the transport coefficients. Considering only up to order $n=1$ in the Hermite series, we have the explicit expressions for the linear case
\begin{align} \label{saturation_length}
 \psi_0 &= \frac{b\nu_0 (1-\alpha_1)\lambda e^{\lambda^2/2}}{1+(1-\alpha_1)(1+\lambda^2)\hat\tau e^{\lambda^2/2}},\\
  L_0 &=\frac{V\tau}{\sqrt{1+(1-\alpha_1)(1+\lambda^2)\hat\tau e^{\lambda^2/2}}}.     
\end{align}
Both the full numerical solution and the approximate analytical expressions indicate that the response amplitude decays with memory while that the smoothing length grows with memory. Indeed, memory makes that the agents \textit{miss} the precise position of the chemotactic signal because, even if the reach an optimal position, they continue moving persistently with reduced tumbling rate in a \textit{wrong} direction. 
As expected, $\psi_0$ grows with the sensitivity $\lambda$, but the dependence of the smoothing length with $\lambda$ is weaker.

To exemplify the character of the response and compare with agent-based simulations, we consider the simple case of a stationary signal with a step profile   in one direction, $l(x) = l_0 + l_1\operatorname{sgn}(x)$, with $\operatorname{sgn}(x) \equiv x/|x|$ the sign function. Here,  the stationary bacterial  density and CheY-P concentration can be explicitly obtained by calculating the inverse  Fourier transform in Eq.~\eqref{eq.responselonretz}. Note that although the response functions [Eq.~\eqref{eq.responselonretz}] are valid for small $k$, the non-local effect decays with $k$ and the Fourier transform converges for the imposed step function, giving
\begin{align} \label{rho_for_step_ligand}
\rho(x)&=\rho_0+ \rho_0  l_1 \operatorname{sgn}(x) \left[\psi_1+\psi_0(1-e^{-k_0|x|})\right], \\
\rho_X(x)&=-\rho_0 \frac{D_{11}}{D_{12}}  \psi_0  l_1 \operatorname{sgn}(x) e^{-k_0\left|  x\right|} . \label{rhoX_for_step_ligand}
\end{align}
Alternatively, these expressions can also be obtained by directly solving the linearized hydrodynamic equations in real space. We recall that the solution of the hydrodynamic equations in real or Fourier space, or even their numerical solution, is much simpler than to directly deal with the kinetic equation for the same geometry as in Ref.~\cite{villa2023kinetic}.
To compare, the Keller--Segel model gives for the density, the local response $\rho(x)=\rho_0 e^{\mu l(x)/D}\approx \rho_0+\rho_0 \psi_0 l_1  \operatorname{sgn}(x)$, where the linear response was given in the second expression.

\begin{figure}[t!]
\begin{center}
     \includegraphics[width=\columnwidth]{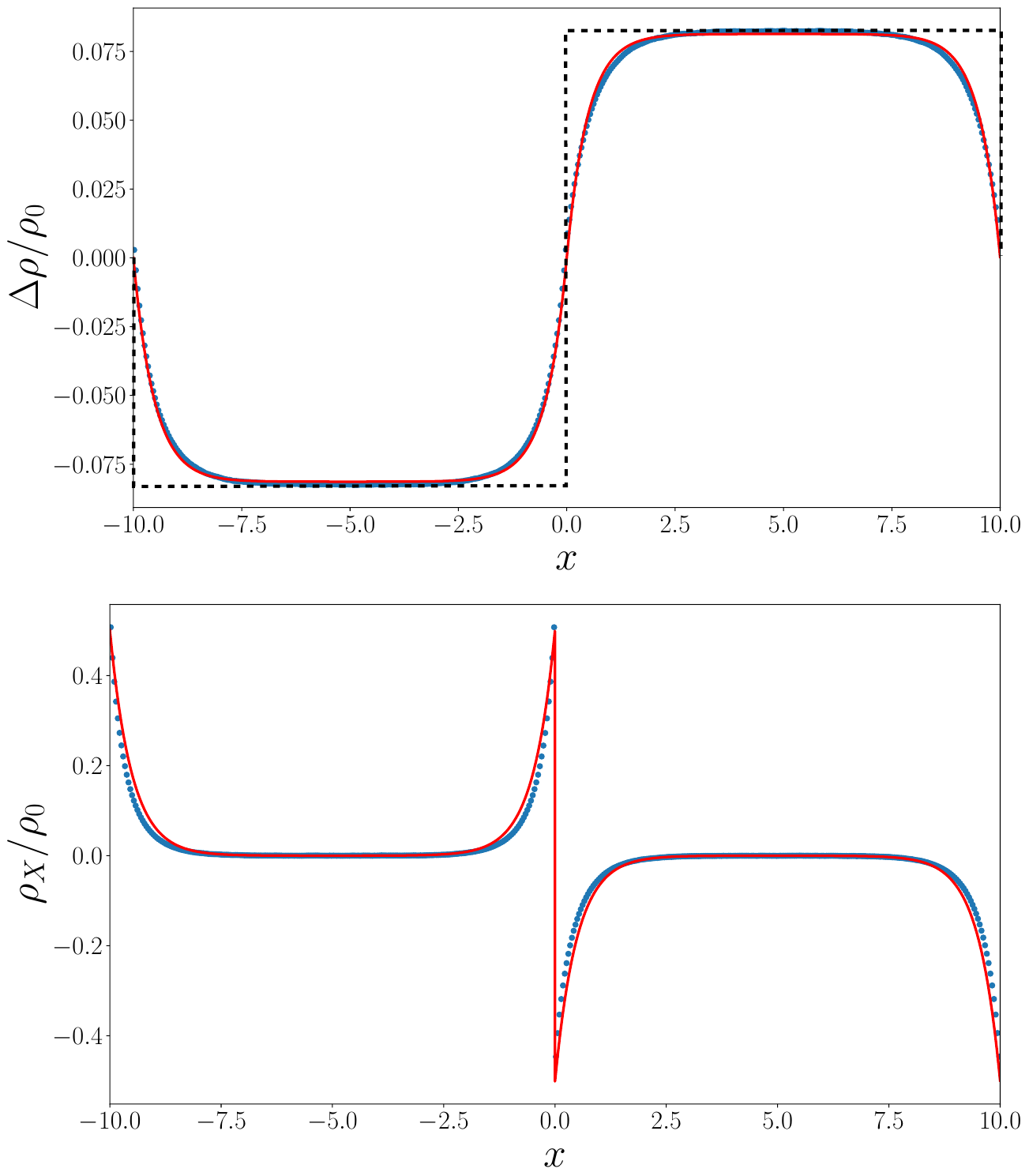}
\end{center}
\caption{Stationary normalized density (top) and protein  density (bottom)  profiles generated for a step function signal in the linear model. The blue circles represent the results of the simulations of particles with $\hat\tau=1.0$, $\lambda=1.0$. In all cases, the particles are confined in a square box of size $L = 20$ with periodic boundary conditions, and with a chemotactic pulse of amplitude $l_1 = 0.4$. For these parameters $L_0=0.43$. The red solid lines are the theoretical predictions and the solid dashed line the prediction of the KS model. The case of fully isotropic tumbling kernel, with $\alpha_1=0$, has been considered. Units have been fixed to $V=\nu_0=b=1$.  }
\label{rho_simulation}
\end{figure}

To verify the analytical predictions, agent-based simulations are performed in two dimensions for the linear model: bacteria move in two spatial dimensions and the directors are characterized by a single angle, which, at tumbles, is sorted uniformly in the interval $[0,2\pi]$. That is, the tumbling kernel is isotropic, implying $\alpha_1=0$. These simulations involve the numerical solution of the stochastic equation~\eqref{langevin_eq_lineal} for $X$  and the position vector $\VEC{r}\equiv x\HAT{x}+y\HAT{y}$ using the Euler--Heun scheme, following the method used in Ref.~\citep{villa2023kinetic} to reproduce the dynamics. The system is initialized homogeneously and, after a  transient to reach the steady state (longer than the memory time $\tau$ and the diffusion time $L^2/D$, where $L$ is the box size), $\rho(x)$, $\rho_X(x)$, $\langle X \rangle(x)$, and  $\langle \nu \rangle(x)$ are measured. The latter is directly measured counting the number of tumbles on $x$ bins. Figure \ref{rho_simulation} presents the comparison between the simulation results and the theoretical predictions for the normalized  bacterial excess density $\Delta\rho /\rho_0 \equiv [\rho(x)-\rho_0]/\rho_0$, and the normalized protein density  $\rho_X(x)/ \rho_0$, showing a very good agreement. 

Both in the theoretical and numerical results, the bacterial density displays a rounded profile, compared with the step response of the Keller--Segel model, and $\rho_X$ presents a sharp discontinuity. At the hydrodynamic level, the steady state is characterized by $\VEC J=0$. The Keller--Segel model [Eq.~\eqref{eq.ks2}] implies that $\rho$ becomes flat as soon as $l$ is uniform, giving rise to the discontinuous density profile. Considering memory, a vanishing mass flux gives
\begin{align}
D_{11}\nabla\left[\left(1-D_{12}\langle X\rangle/D_{11}\right)\rho\right] = \mu_{11}\rho_0\nabla l,
\end{align}
where the linearized form of Eq.~\eqref{eqJJxfinal.J} was used. Then, over the distance where $\langle X\rangle\equiv \rho_X/\rho$ has not relaxed,  $\rho$ will not be uniform even if $l$ is uniform. This is precisely the smoothing distance $L_0$.
This smoothing length can be well understood also in terms of the microscopic chemotactic model. For concreteness, consider a bacterium that crosses the boundary from the poor to the rich ligand region. In this case, $l$ experiences an abrupt increase, meaning that $\dot l$ is characterized by a Dirac delta function with positive weight. As a consequence, $X$ suffers an instantaneous decrease as given by Eq.~\eqref{langevin_eq_general}. This abrupt change in $X$ manifests as a discontinuity in $\rho_X$. After that, as the ligand does not change, $X$ will relax on a time $\tau$ to zero, which gives a length scale of the order of $L_0\sim V\tau$. During this relaxation time, the tumbling rate is smaller than average, implying that bacteria moving from the poor to the rich ligand region tumble less than those moving in the opposite direction. This imbalance, on the length scale $L_0$, produces a net bacterial flux toward the ligand rich region that will stop once the resulting excess density generates a diffusive flux in the opposite direction, such as that the total bacterial flux vanishes. The final density profile will therefore be smoothed over the same length scale.

\begin{figure}[t!]
\begin{center}
     \includegraphics[width=\columnwidth]{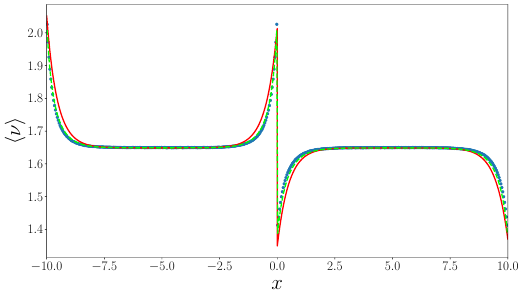}
\end{center}
\caption{ Tumble rate average profile for the case described in Fig.~\ref{rho_simulation}.
The blue circles represent the results of the simulations, the red solid line is the theoretical prediction Eq.~\eqref{eq:numean1} by using $\langle X \rangle$ given by Eq.~\eqref{eq:mean_x}, and the green dashed line is the theoretical prediction  \eqref{eq:numean2} using the measured values of $\rho(x)$ and $\rho_X(x)$.  }
\label{numean_simulation}
\end{figure}

Experimentally, the bacterial density is measurable, but $\rho_X$ can be difficult to obtain. A related observable, which is accessible to experiments \cite{khan2004fast,junot2022run} is the position-dependent average tumbling rate $\langle \nu\rangle(\VEC{r})$, which in the linear model equals $\nu_0 \langle e^{\lambda X}\rangle(\VEC{r})$. To obtain it in the present formalism,  we use the cumulant-generating function \cite{van1992stochastic}, which for the random variable $X$ with parameter $\lambda$ can be defined as
$K(\lambda) =\operatorname{ln}\left( \langle e^{\lambda X } \rangle \right)$.
Expanding this expression in Taylor series, we obtain
\begin{align} \label{cumulant_taylor_expantion}
K(\lambda) =\sum_{n=1}^\infty \kappa_n \frac{\lambda^n}{n!} = \lambda \langle X \rangle + \frac{\lambda^2}{2} \sigma_X^2 + \cdots,
\end{align}
where the omitted terms depend on higher cumulants $\kappa_n$ of $X$.
Taking into account that in the linear model $X$ is well described by a normalized  Gaussian variable, i.e. the only nonzero cumulants are the first and the second, where the  standard deviation is  $\sigma_X = 1$, it results $\operatorname{ln}\left( \langle \nu/\nu_0 \rangle (\VEC{r}) \right) = \lambda \langle X \rangle(\VEC{r}) + \frac{\lambda^2}{2}$.
Finally,  we obtain for the average tumbling rate
\begin{subequations}
\begin{align}
  \langle\nu \rangle (\VEC{r}) &=   \left( \nu_0 e^{\lambda^2/2} \right) e^{\lambda \langle X\rangle(\VEC{r})}, \label{eq:numean1}\\
  &=  \left( \nu_0 e^{\lambda^2/2} \right)e^{\lambda \rho_X(\VEC{r})/\rho(\VEC{r})}, \label{eq:numean2}
\end{align}
\end{subequations}
where we used  that $\langle X\rangle=\rho_X/\rho$.
 In the case of ligand with a step profile,  
\begin{align} \label{eq:mean_x}
\langle X \rangle = \frac{ D_{11} l_1 \mu_{11} \text{sgn}(x)}{D_{12} D_{11} -D_{12} l_1\mu_{11}  \text{sgn}(x)(1-e^{k_0\left| x\right| })}.
\end{align} 
The average tumbling rate is displayed in Fig.~\ref{numean_simulation}, where we compare the values obtained in the simulation  with the expression \eqref{eq:numean1} by using the explicit expression of  $\langle X \rangle$ given by Eq. \eqref{eq:mean_x}, and with the expression \eqref{eq:numean2} using the measured values of $\rho(x)$ and $\rho_X(x)$.    The agreement between the simulations and the theoretical results is good, with better agreement using the simulation measures of $\langle X \rangle$.

\subsection{Traveling wave}
\begin{figure}[t!]
\begin{center}
\includegraphics[width=.9\columnwidth]{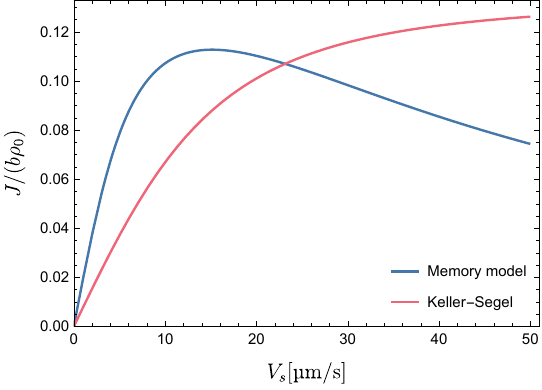}
\end{center}
\caption{Normalized bacterial current as a response to a traveling chemotactic wave of speed $V_s$, computed using the values of the linear model for \textit{E.\ coli}. The current is computed for a periodic box of length $L=\SI{800}{\micro\meter}$, equal to the value in the experiment~\cite{li2017barrier}, meaning that $\Psi_\rho$ is evaluated at $k=2\pi/L$. In blue the prediction of this article and in red the prediction of the  Keller--Segel theory (divided by 5 to help the comparison).}
\label{fig.current}
\end{figure}

For a more complex statio-temporal response, here we consider the case of a traveling chemotactic wave $l(x,t)=l_0 e^{i k(x-V_st)}$. In this case, the response is obtained by substituting $\omega=Vk$ in $\Psi$ and $\Psi_X$ given in Appendix~\ref{expressions.gral}. Replacing in  Eqs.~\eqref{complet_rho_eq} and \eqref{eqJJxfinal.J}, gives for the linear response of the chemotactic current 
\begin{align}
J = V_s \Psi_\rho(k,V_sk) l_0,
\end{align}
while in the Keller--Segel theory, the response is
\begin{align}
J_\text{KS} = \frac{k \mu V_s}{Dk^2-i k V_s} l_0.
\end{align}
Recent experiments performed with \textit{E.\ coli}, showed that the resulting chemotactic current is not monotonic with the wave speed $V_s$, presenting a maximum at $V_s\approx\SI{8}{\micro\meter/\second}$~\cite{li2017barrier}. Figure~\ref{fig.current} shows the predicted current using the linear model with the fitted values for \textit{E.\ coli} (see Section~\ref{sec.ecoli}), which is compared with the prediction of the  Keller--Segel theory.  Although the experiment is performed in a strong non-linear response regime, where the current saturates to large values, it is remarkable that the present model predicts the existence of the maximum for a wave velocity in the same order or magnitude, while the Keller--Segel theory gives a monotonically increasing current failing event to predict the existence of a maximum. 
These results are consistent with the analysis in Ref.~\cite{li2017barrier}, where it is shown that the origin of the maximum is the existence of a finite memory. 
 
\subsection{Diffusive dynamics}

As a final case of interest,  we  study the bacterial dynamics in absence of a chemotactic signal. In this case, Eqs.~\eqref{rho_eq} and \eqref{rhox_eq}, with the transport relations Eqs.~\eqref{eq8} and \eqref{eq14}, reduce to
\begin{align} \label{matrix_eqs_for_rho_and_rho_x_fourier}
   \pdv{}{t} \begin{pmatrix}
        \tilde{\rho} \\
        \tilde{\rho}_X 
    \end{pmatrix} = -  M  \begin{pmatrix}
        \tilde{\rho} \\
        \tilde{\rho}_X 
    \end{pmatrix}.
\end{align}
for the Fourier modes 
\begin{subequations}\label{eq:matrix_form_adim_t}
\begin{align} 
 \tilde{\rho} \equiv &   \int d \VEC{r}  \rho e^{i \VEC{k} \cdot \VEC{r}},\\ 
 \tilde{\rho}_{X }  \equiv &    \int d \VEC{r}  \rho_X e^{i \VEC{k} \cdot \VEC{r}}.
\end{align} 
\end{subequations}

The dynamic matrix $M$ is 
\begin{align} \label{matrix_L}
M \equiv  
\begin{pmatrix}
        D_{11}k^2  & - D_{12} k^2 \\
        - D_{12} k^2  &  D_{22} k^2 +\frac{\gamma_1}{ \tau }   
    \end{pmatrix},
\end{align}
where we used that $D_{21}=D_{12}$.
This matrix is symmetric, has real coefficients and is semidefinite positive, implying that it is diagonalizable with positive and real eigenvalues $\chi_i$, and eigenvectors  $\VEC{v}_i$,   with $i=1,2$. Then, the general solution is of the form
\begin{align}\label{eq:sol}
\begin{pmatrix}
 \tilde{\rho} \\
        \tilde{\rho}_X 
    \end{pmatrix}  = \sum_{j=1}^2 c_i \VEC{v}_i  e^{-  \chi_i t },
\end{align}    
with the constants $c_i$ determined by the initial conditions. 

\begin{figure}[htb]
    \centering
    \includegraphics[width=\columnwidth]{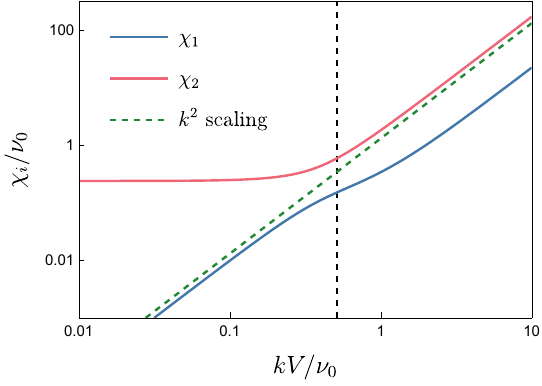}
    \caption{Relaxation eigenvalues $\chi_i$, $i=1,2$ , as a function of $k$ in log-log scale for the linear model. The blue and red  solid lines correspond to $\chi_1$ and $\chi_2$, respectively. The green dashed line represents $k^2$, in order to identify the  modes that presents diffusive behavior in the different regimes. The vertical dashed line corresponds to $k^* \equiv (D_{22} \tau)^{-1/2}$, separating the different dynamical regimes.  The values used are $\lambda =1.62$, $\hat\tau = 4.2 $, and $\alpha_1=0.33$.}
    \label{fig:eigenvalues}
\end{figure}

   Figure \ref{fig:eigenvalues} shows  the eigenvalues $\chi_{1,2}$ as a function of $k$ in logarithmic scale for the lineal model. The values of the sensitivity to fluctuations and the memory time used for the plot are $\lambda=1.62$ and $\hat\tau =4.2$, corresponding to the experimental values of  \textit{E.\ coli} \cite{figueroa20203d}. For small $k$, there is  one eigenvalue that saturates to a constant and another one that goes as $k^2$, reflecting a diffusive mode. For large $k$, now the two modes are diffusive, with new diffusion constants, one of those being smaller than the small wavevector diffusivity. To describe in more detail this behavior and determine the crossover wavelength $k^*$, we analyze separately the cases of small and large wavevectors.

Expanding the eigenvalues in a Taylor series in $k$, and keeping terms up to $k^2$, we get
\begin{align}
   \chi_1 = &   D_{11} k ^2  +\mathcal{O}(k^4) ,  \label{eq:L_eigenvalues1} \\ 
   \chi_2 = & \frac{\gamma_1}{\tau}  +D_{22} k^2+ \mathcal{O}(k^4), \label{eq:L_eigenvalues2}
\end{align}
with their corresponding eigenvectors, truncated to leading order in $k$,
\begin{align}
   \VEC{v}_1 = &    \begin{pmatrix}
   1\\
   \tau D_{12} k^2/\gamma_1
\end{pmatrix}   ,    \label{eq:L_eigenvectors1} \\    
   \VEC{v}_2 = &  - \begin{pmatrix}
       \tau  D_{12} k^2/\gamma_1  \\
  - 1
\end{pmatrix} .  \label{eq:L_eigenvectors2}
\end{align}
Consistent with the figure, there is only one diffusive mode with diffusion coefficient $D_{11}$, associated with the eigenvector $\VEC{v}_1$, which is essentially the density mode. As discussed in Sec.~\ref{sec.longvsshort}, this diffusion coefficient equals the short memory value $D$.
The other mode, which is associated to $\rho_X$, relaxes at a constant rate $\tau$ in the limit of vanishing wavevectors. Note that in this regime, $\rho$ and $\rho_X$ are practically decoupled.

\begin{figure*}[htb]
\includegraphics[width=2\columnwidth]{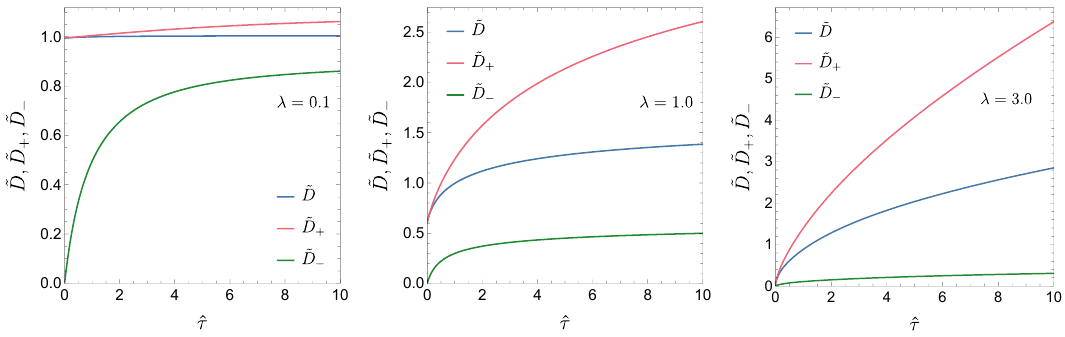}
\caption{Small $k$ dimensionless diffusion coefficient $\tilde D=d\nu_0 D/V^2$ (blue), and large $k$ dimensionless diffusion coefficients $\tilde D_+ = d\nu_0 D_+/V^2$ (red) and $\tilde D_- = d\nu_0 D_-/V^2$ (green) for the linear model as a function of the dimensionless memory $\hat\tau$   for selected values of the sensitivity $\lambda$. 
The solid lines correspond evaluating the transport coefficients using the numerical solution of the system of equations \eqref{eqsalgM-Q} truncating it  up to $n=10$. In all plots, we consider a fully isotropic tumbling kernel, with $\alpha_1=0$.}
\label{fig.Dplusminus}
\end{figure*}

To study the regime of high $k$, we take the limit $1/\tau \to 0 $ in the general expression for $\chi_1$ and $\chi_2$, obtaining
\begin{align}
   \lim_{1/\tau \to 0 }\chi_{1,2} =& \left(\frac{D_{11}+D_{22}\mp \sqrt{(D_{11}-D_{22})^2+4 D_{12}^2}}{2}\right)k^2.
\end{align}
The modes are indeed diffusive, with diffusion coefficients $D_\mp$ given by the term inside the parenthesis. It is direct to verify that $D_-<D<D_+$, as was identified in the figure. Figure~\ref{fig.Dplusminus} displays these three diffusion coefficients for the linear model as a function of the model parameters, where it can be verified that $D_-$ can be notoriously smaller than $D$, implying long relaxation times. In this regime, the eigenvectors (not written by directly obtained from the dynamic matrix) couple $\rho$ and $\rho_X$ at the same order, so both fields evolves diffusively. 

The crossover between the two regimes can be directly obtained by observation of the matrix [Eq.~\eqref{matrix_L}] or from Eq.~\eqref{eq:L_eigenvalues2}, resulting in $k^* \equiv (D_{22} \tau/\gamma_1)^{-1/2}$. In the case of large memory times, the associated crossover wavelength is  also  large. 

\section{Numerical values for  \textit{E.\ coli}}  \label{sec.ecoli}
The parameters of the linear model for \textit{E.\ coli} (RP437 bacteria in motility buffer supplemented with serine) were determined in Ref.~\cite{figueroa20203d}: $V = \SI{27}{\micro\meter/\second}$, $\nu_0 = \SI{0.22}{\second^{-1}}$, $ \tau = \SI{19}{\second}$, and $\lambda = 1.62$. Also, the first moment of the tumble kernel is known: $\alpha_1\approx0.33$ \cite{berg1972chemotaxis}. The only remaining parameter is $b$, which depends on the specific ligand to be considered. 

With these results, it is possible to provide explicit values for the transport coefficients and related parameters discussed in previous sections. 
The diffusion coefficients are $D_{11}=\SI{1.3E3}{\micro\meter^2/\second}$,  $D_{12}=D_{21}=\SI{0.81E3}{\micro\meter^2/\second}$,  and  $D_{22}=\SI{0.99E3}{\micro\meter^2/\second}$. The chemotactic mobilities are $\mu_{11}/b = \SI{0.42E2}{\micro\meter^2/\second^2}$, $\mu_{12}/b = \SI{2.2E2}{\micro\meter^2/\second^2}$, $\mu_{21}/b = \SI{0.52E2}{\micro\meter^2/\second^2}$, and $\mu_{22}/b = \SI{2.6E2}{\micro\meter^2/\second^2}$. The amplitude and smoothing length of the static response function are $\psi_0/b=\SI{0.032}{\second^{-1}}$ and $L_0=\SI{1.7E2}{\micro\meter}$, respectively. Finally, the large wavevector diffusivities are $D_+ = \SI{2.0E3}{\micro\meter^2/\second}$ and $D_- = \SI{0.33E3}{\micro\meter^2/\second}$, which are valid for wavelengths smaller than $L^*=2\pi/k^*=\SI{1.5E3}{\micro\meter}$.

\section{Conclusions} \label{sec.conclusions}
The bacterial chemotactic response can be rationalized in terms of a small number of variables associated to the internal concentration of relevant proteins inside the bacterial body, which control the tumble rate. In the case of \textit{E.\ coli}, there is a strong scale separation between the relaxation rates of these proteins allowing for a reduction of the model to a single variable, the concentration of the CheY-P protein governed by a single memory time $\tau$. Inspired by this model, but considering general non-linear coupling and responses to the ligand, it is possible to write down a kinetic equation for an ensemble of bacteria in presence of a ligand field (food, chemoattractant, or chemorepellent), with arbitrary spatio-temporal dependence [Eq.~\eqref{kinetic_eq}]. This kinetic equation considers the evolution of an effective internal variable $X$ that govern the chemotactic process, which we assume is controlled by a single large memory time $\tau$, while all other characteristic times are shorter. In the simpler model, $X$ is the CheY-P protein concentration inside the bacterial body, $\tau$ is the methylation time of the chemorecetors, and  coupling and responses to the ligand are assumed to be simple linear functions of $X$. This linear model is analyzed throughout the article as it allows to provide explicit results and because its parameters have been measured for  \textit{E.\ coli}.

The kinetic analysis shows that there is a emergent long length scale, which in the case of \textit{E.\ coli} can be of several hundreds of micrometers, making it necessary to build a practical framework when the ligand field varies on these scales as it happens in natural an artificial microfluidic environments.
Employing the Chapman--Enskog method, we derived the hydrodynamic equations that describe a bacterial ensemble as a formal expansion in spatial gradients. In the case the memory time is small, the methods yields to the standard Keller--Segel equation for the bacterial density $\rho$, with a chemotactic mobility and diffusion coefficient computed entirely in terms of the microscopic parameters [Eqs.~(\ref{eq7}-\ref{eq10})]. More  relevant is the case of long memory time, as it has been determined experimentally for \textit{E.\ coli}. In this case, besides the bacterial density, the density of the internal variable $\rho_X$ emerges as a new relevant field, and the coupled equations for $\rho$ and $\rho_X$ are derived. They have the form of reaction-diffusion equations, which are coupled to the ligand field [Eqs.~(\ref{complet_rho_eq}-\ref{eqs.constlaws})]. Associated to these equations, the diffusion and mobility transport coefficients are obtained in terms of the microscopic parameters.

The derived equations are analyzed in simple regimes, aiming to highlight some relevant features. In the case of a uniform and stationary ligand gradient, a stationary chemotactic current is obtained. However, the coupling with $\rho_X$ generates for a transient $\tau$, an effective chemotactic mobility larger than the stationary one if the ensemble was previously placed in a region with a strong ligand gradient. Also, analyzing the linear response for signals with spatio-temporal dependence, a nonlocal response in time and space is obtained, which is absent in the Keller--Segel description. The associated smoothing length equals \SI{170}{\micro\meter} for the case of \textit{E.\ coli}, which is comparable to many spatial features in natural and artificial microfluidic environments. Note that rotational diffusion could reduce slightly the smoothing length. Experimentally, for the linear model, the nonlocal response can be determined by measuring the density profiles and the local tumbling rate when the ensemble is placed in presence of an inhomogeneous signal. Appropriate observables should be defined for the case of other chemotactic models. 
 When analyzing the response to chemoattractant traveling waves, the response current presents a maximum for a particular wave velocity as has been observed in experiments~\cite{li2017barrier}, maximum that is absent in the Keller--Segel description.  Finally, the diffusive dynamics in absence of signal is analyzed, showing that the diffusion coefficient is scale dependent, with a crossover at \SI{1.5E3}{\micro\meter}. 
Besides the cases considered in this article, these equations can be used is different relevant experimental or numerical configurations. 

The derivation of the hydrodynamic equations and the transport coefficients has been presented in detail for future use, in case other chemotactic and motility models need to be considered, for example, dealing with saturation of the ligand receptors, more complex chemotactic circuits, or bacterial variability (see for example, \cite{tu2008modeling,tu2013quantitative}). Possible extensions to the model analyzed in this manuscript can include the modification of the tumbling kernel with ligand gradient~\cite{vladimirov2010predicted,saragosti2011directional,saragosti2012modeling}, or chemokinesis, in which bacteria modify the propulsion speed  as a response to ligand concentration~\cite{armitage1997bacterial,garren2014bacterial,garren2014bacterial}. In some cases, only the transport coefficients will change but in others, for example, when dealing with bacterial interactions, also the hydrodynamic equations could be modified.

The hydrodynamic equations derived in this manuscript need to be complemented with appropriate boundary conditions. A first approach would be simply to impose non-flux boundary conditions for both fields. However, the problem is far from trivial because bacteria and other  microswimmers, either by simple persistence of motion or by hydrodynamic attraction, tend to accumulate on surfaces~\cite{berke2008hydrodynamic}, where they modify their motion~\cite{lauga2006swimming}. More research is needed to describe  the interaction of bacteria with surfaces, considering memory effects.

In Refs.~\cite{hein2016physical,jakuszeit2021migration} the case of bacteria responding to diffusing point sources taking place in natural environments is analyzed, showing that due to the spatio-temporal dependence of the signal, non-trivial responses can emerge. This and similar configurations are perfect examples where the equations derived here can be applied, either numerically or analytically. A proper analysis, considering different regimes goes beyond the scope of the manuscript and will be devoted to future research.

\begin{acknowledgments}

This research is supported by Fondecyt Grants No.~1220536 (RS) and ANID Millennium Science Initiative Program NCN19$\_$170, Chile. MM acknowledges financial support by grant ProyExcel\_00505 funded by
Junta de Andalucía and grant PID2021-126348NB-I00 funded by
MCIN/AEI/10.13039/501100011033/ and ERDF ``A way of making Europe''.
\end{acknowledgments}

\appendix

\begin{widetext}

\section{Analytical solution for the linear model in the long memory case} \label{app.solutionlinear}

Keeping up to $n=1$, that is, considering two polynomials, the solution of Eqs.~\eqref{eqsalgM-Q} is
\begin{subequations}
\begin{align} \label{3.46}
O_0  =& -\frac{1}{e^{\frac{\lambda^2}{2}}}\left[\frac{1}{1-\alpha_1}+\frac{\lambda^2 \hat\tau e^{\frac{\lambda^2}{2}}}{1+(1-\alpha_1)e^{\frac{\lambda^2}{2}} \hat\tau}\right],\\
O_1  =& \frac{\lambda \hat\tau}{1+(1-\alpha_1) e^{\frac{\lambda^2}{2}} \hat\tau},\\
P_0 =& Q_0/b  = \frac{\lambda \hat\tau}{1+(1-\alpha_1) e^{\frac{\lambda^2}{2}} \hat\tau},\\
P_1  =& Q_1/b  = -\frac{\hat\tau}{ 1+(1-\alpha_1) e^{\frac{\lambda^2}{2}} \hat\tau},\\
R_0 =& R_1 =0.
\end{align}
\end{subequations}
where the expression \eqref{matrixbmm_linear} for the $b_{mm'}$ matrix was used.

If we truncate up to $n=2$,  that is, considering three polynomials, the results are
\begin{subequations}
\label{eqM.orden2}
\begin{align} \label{3.47}
                  O_0  =&-\frac{e^{-\frac{\lambda^2}{2}} \left[(1-\alpha_1)^2 e^{\lambda^2} \left(\lambda^4+2 \lambda^2+2\right) \hat\tau^2+(1-\alpha_1) e^{\frac{\lambda^2}{2}} \left(\lambda^4+8 \lambda^2+6\right) \hat\tau+4\right]}{2(1-\alpha_1) \left[(1-\alpha_1)^2 e^{\lambda^2} \hat\tau^2+(1-\alpha_1) e^{\frac{\lambda^2}{2}} \left(2 \lambda^2+3\right) \hat\tau+2\right]}, & \\ 
        O_1  = &{\frac{\lambda \hat\tau \left[ (1-\alpha_1)e^{\frac{\lambda^2}{2}} \left(\lambda^2+1\right) \hat\tau+2\right]}{\left[(1-\alpha_1)^2e^{\lambda^2} \hat\tau^2+(1-\alpha_1)e^{\frac{\lambda^2}{2}} \left(2 \lambda^2+3\right) \hat\tau+2\right]}},\\
        O_2  =&-{\frac{\lambda^2 \hat\tau \left[(1-\alpha_1)e^{\frac{\lambda^2}{2}} \hat\tau-1\right]}{\sqrt{2}\left[(1-\alpha_1)^2e^{\lambda^2} \hat\tau^2+(1-\alpha_1)e^{\frac{\lambda^2}{2}} \left(2 \lambda^2+3\right) \hat\tau+2\right]}},
\end{align}
\end{subequations}
\begin{subequations}
\label{eqN.orden2}
\begin{align}
                  P_0  =Q_0/b = &{\frac{\lambda \hat\tau \left[(1-\alpha_1)e^{\frac{\lambda^2}{2}} \left(\lambda^2+1\right) \hat\tau+2\right]}{(1-\alpha_1)^2e^{\lambda^2} \hat\tau^2+(1-\alpha_1)e^{\frac{\lambda^2}{2}} \left(2 \lambda^2+3\right) \hat\tau+2}}, \\ 
        P_1  =Q_1/b = &-{\frac{\hat\tau \left[(1-\alpha_1)e^{\frac{\lambda^2}{2}} \left(2 \lambda^2+1\right) \hat\tau+2\right]}{ \left[(1-\alpha_1)^2e^{\lambda^2} \hat\tau^2+(1-\alpha_1)e^{\frac{\lambda^2}{2}} \left(2 \lambda^2+3\right) \hat\tau+2\right]} },\\
        P_2  =Q_2/b =& {\frac{\sqrt{2}(1-\alpha_1)e^{\frac{\lambda^2}{2}} \lambda \hat\tau^2}{(1-\alpha_1)^2 2 e^{\lambda^2} \hat\tau^2+(1-\alpha_1)2 e^{\frac{\lambda^2}{2}} \left(2 \lambda^2+3\right) \hat\tau+4}},
    \end{align}
\end{subequations}

\begin{subequations}
\label{eqP.orden2}
\begin{align}                  
{R_0/b  =} &\frac{2 (1-\alpha_1 ) \lambda ^2 e^{\frac{\lambda ^2}{2}} \hat\tau ^2}{2 \left(\lambda ^2+1\right)+(1-\alpha_1 ) e^{\frac{\lambda ^2}{2}} \left(\lambda ^2+2\right) \lambda ^2 \hat\tau -2 (1-\alpha_1 )^2 e^{\lambda ^2} \hat\tau ^2}  , \\ 
       { R_1/b  =} &-\frac{ 2\lambda  \hat\tau  \left[\lambda ^2+2-2 (1-\alpha_1) e^{\frac{\lambda ^2}{2}} \hat\tau \right]}{2 (1-\alpha_1)^2 e^{\lambda ^2} \hat\tau ^2-(1-\alpha_1 ) e^{\frac{\lambda ^2}{2}} \left(\lambda ^2+2\right) \lambda ^2 \hat\tau -2 \left(\lambda ^2+1\right)},\\
       { R_2/b  =}& \frac{2\sqrt{2}\hat\tau  \left[\lambda ^2+1-(1-\alpha_1 ) e^{\frac{\lambda ^2}{2}} \hat\tau \right]}{2 (1-\alpha_1 )^2 e^{\lambda ^2} \hat\tau ^2-(1-\alpha_1 ) e^{\frac{\lambda ^2}{2}} \left(\lambda ^2+2\right) \lambda ^2 \hat\tau -2 \left(\lambda ^2+1\right)} .
    \end{align}
    \end{subequations}

\section{Small memory time limit for the transport coefficients in $\VEC{J}$} \label{app.limits}

First, we recall that the operator $\mathcal{\widehat F}$ is not invertible, and its kernel is $\phi=U_0$. Then, the operator 
\begin{align}
\mathcal{\widehat  L}_0^\prime &= \mathcal{\widehat  F} - \nu_0\tau (1-\alpha_1)C(X) \nonumber\\
&=\frac{\partial^2}{\partial X^2}+\frac{\partial}{\partial X} A(X)- \nu_0\tau (1-\alpha_1)C(X)
\end{align}
becomes also non invertible and with the same kernel when $\nu_0\tau\to0$. However, in Eq.~\eqref{c} for $P$, the RHS is proportional to $U_1$, which is orthogonal to the kernel, and therefore the solution of this equation is proportional to  $\nu_0\tau$. As a consequence $D_{12}$ vanishes in this limit by Eq.~\eqref{eq9.2A}.
In Eq.~\eqref{eq.paraR}, $\gamma_1$ is strictly positive making the total operator invertible. As a consequence, $R$ is proportional to  $\nu_0\tau$ and the associated transport coefficient $\mu_{12}$ vanishes in the limit of small memory by Eq.~\eqref{eq9.4A}.

On the other hand, in Eq.~\eqref{3.31} for $O$, the RHS belongs to the kernel of $\mathcal{\widehat F}$. Therefore, in the limit $\nu_0\tau\to0$ both the LHS and the RHS go to zero simultaneously, resulting in a solution that is of order 1. By Eq.~\eqref{eq9.1A}, this implies that $D_{11}$ remains finite in the limit. Finally, in Eq.~\eqref{3.33b} for $Q$, the RHS is not orthogonal to the kernel of the singular operator, implying that the solution remains finite in the limit, and so it does $\mu_{11}$ by Eq.~\eqref{eq9.3A}.

\section{Linear response functions}\label{expressions.gral}
\begin{subequations}
\begin{align}
\Psi_\rho(k,\omega) &= \frac{k^2 [\gamma_1 \mu_{11} + k^2 (D_{22} \mu_{11} - D_{12} \mu_{21}) \tau - 
   i (D_{12} g_1 + \mu_{11}) \tau \omega]}{
D_{11} \gamma_1 k^2 + (-D_{12}^2 + D_{11} D_{22}) k^4 \tau - 
 i [\gamma_1 + (D_{11} + D_{22}) k^2 \tau] \omega - \tau \omega^2},\\
 \Psi_X(k,\omega)&= \frac{\tau [-D_{12} k^4 \mu_{11} + (D_{11} k^2 - i w) (k^2 \mu_{21} + 
      i g_1 \omega)]}{-D_{11} \gamma_1 k^2 + (D_{12}^2 - D_{11} D_{22}) k^4 \tau + 
 i [\gamma_1 + (D_{11} + D_{22}) k^2 \tau] \omega + \tau \omega^2}.
\end{align}
\end{subequations}

\end{widetext}

\end{document}